\documentclass[epj,final]{svjour}
\usepackage{amsmath} 
\usepackage{widetext}
\usepackage{graphicx} 
\usepackage{slashed} 
\usepackage{hyperref}
\usepackage[numbers,sort&compress]{natbib}
\allowdisplaybreaks
\def\Feynarts{{{\sc FeynArts}}}
\def\Feyncalc{{{\sc FeynCalc}}}
\def\PackageX{{{\sc Package-X}}}
\def\Mathematica{{{\sc Mathematica}}}
\def\beq{\begin{equation}}
\def\eeq{\end{equation}}
\def\beqn{\begin{eqnarray}} 
\def\eeqn{\end{eqnarray}}
\def\nn{\nonumber}
\def\Eq#1{Eq.~(\ref{#1})}
\def\uv{{\rm UV}}
\newcommand{\valencia}{Instituto de F\'{\i}sica Corpuscular, Universitat de Val\`{e}ncia -- Consejo Superior de Investigaciones Cient\'{\i}ficas, \\
Parc Cient\'{\i}fic, E-46980 Paterna, Valencia, Spain.}

\begin{document}

\title{On the interplay between the loop-tree duality and helicity amplitudes}
\author{F\'elix Driencourt-Mangin\inst{1}\thanks{e-mail: \href{mailto:felix.dm@ific.uv.es}{felix.dm@ific.uv.es}}
 \and Germ\'an Rodrigo\inst{1}\thanks{e-mail: \href{mailto:german.rodrigo@csic.es}{german.rodrigo@csic.es}} 
\and Germ\'an F. R. Sborlini\inst{1}\thanks{e-mail: \href{mailto:german.sborlini@ific.uv.es}{german.sborlini@ific.uv.es}}
\and William~J.~Torres~Bobadilla\inst{1}\thanks{e-mail: \href{mailto:william.torres@ific.uv.es}{william.torres@ific.uv.es}}
}
\institute{\valencia}
\date{\today}
%
\abstract{
The spinor-helicity formalism has proven to be very efficient in the calculation of scattering amplitudes in quantum field theory, while the loop tree duality (LTD) representation of multi-loop integrals exhibits appealing and interesting advantages with respect to other approaches. In view of the most recent developments in LTD, we exploit the synergies with the spinor-helicity formalism to analyse illustrative one- and two-loop scattering processes. We focus our discussion on the local UV renormalisation of IR and UV finite amplitudes and present a fully automated numerical implementation that provides efficient expressions which are integrable directly in four space-time dimensions.
}

\setcounter{page}{1}
\maketitle

\section{Introduction}
\label{sec:introduction}
In order to unveil the composition of matter and its interactions, it is necessary to analyse highly-precise experimental data obtained from colliders using accurate predictions. However, the established theoretical frameworks, i.e. the Standard Model, involves very complicated mathematical equations, whose exact solutions are unknown in many physically relevant processes. Thus, most of the computations performed nowadays rely on the perturbative approach, which naturally leads to the appearance of Feynman amplitudes and loop integrals.

With the purpose of achieving a higher accuracy in the theoretical predictions, it is mandatory to explore higher perturbative orders and compute multi-loop amplitudes with high multiplicity. To tackle these calculations, several methods have been developed in the last years. On one hand, there has been an enormous progress in the algebraic handling of scattering amplitudes in 
gauge theories due to the use of alternative kinematic variables as the ones provided by the
spinor-helicity formalism~\cite{Dixon:1996wi}. Also, there was an important improvement due to the study of the mathematical properties 
of the scattering amplitudes. 
For instance in the colour sector~\cite{Bern:1990ux,Kleiss:1988ne,Bern:2008qj}
and the development of new regularisation strategies~\cite{Gnendiger:2017pys,TorresBobadilla:2020ekr}.\footnote{For a detailed review on analytic and semi-numerical techniques for NNLO, we refer the reader to Ref.~\cite{Heinrich:2020ybq}.}
These techniques lead to a much more efficient treatment of the scattering amplitudes, exploiting several symmetries to simplify the underlying expressions.
On the other hand, there were also great advances in the calculation of multi-loop Feynman integrals, 
both analytically and numerically. In particular, pointing towards a more efficient numerical implementation, 
we have been developing a novel strategy based on the loop-tree duality (LTD) theorem~\cite{Catani:2008xa,Bierenbaum:2010cy,Bierenbaum:2012th,Aguilera-Verdugo:2019kbz,Verdugo:2020kzh,Plenter:2020lop,Aguilera-Verdugo:2020kzc,Ramirez-Uribe:2020hes,Aguilera-Verdugo:2020nrp}. 
This theorem allows to decompose any loop amplitude (or loop integral) as the sum of tree-level like objects integrated over a proper phase-space region. From the physical point of view, loop particles are converted into real-radiation ones. From the mathematical side, the integration domain is transformed from a Minkowski to an Euclidean space. 
In fact, the numerical evaluation of multi-loop integrals through LTD is, w.r.t. the approaches that pass by Feynman parametrisation or Mellin-Barnes transformations, more efficient as the number of integrations to be performed does not scale with the number of external particles. 
Very recently, a novel LTD-inspired representation of multi-loop multi-leg scattering amplitudes was presented in Refs.~\cite{Verdugo:2020kzh,Aguilera-Verdugo:2020kzc,Ramirez-Uribe:2020hes,Aguilera-Verdugo:2020nrp}. This strategy, based on the nested residue strategy,
leads to very compact integrand-level representations which 
are free of unphysical singularities.
Likewise, alternative studies of LTD have been presented in Refs.~\cite{Tomboulis:2017rvd,Runkel:2019yrs,Capatti:2019ypt,Capatti:2020ytd,Capatti:2020xjc,Prisco:2020kyb}. 

In this paper, we study how to apply the LTD formalism to the calculation of multi-loop helicity amplitudes. We exploit the fact that LTD works at the level of denominators and the structure of the numerator does not generate any additional difficulty. To this end, we start considering illustrative examples in which the simplicity of the latter is displayed. In order to generate helicity amplitudes, we make use of the spinor-helicity formalism, where we write definite states for the external wave functions. On top of it, to have a very compact expression for the integrand of the helicity amplitude, we also use the momentum twistors' variables~\cite{Hodges:2009hk}. These variables, due to their mathematical properties, allow us to express any kinematic process in terms of the minimal set of variables. In other words, for a process with $n$ external massless particles we have $3n-10$ invariants to deal with~\cite{Badger:2013gxa,Badger:2016uuq}. 
Likewise, the extension to massive particles is straightforward. 

Besides the clearness LTD offers us to compute any multi-loop amplitude, in this paper, we also want to stress on the local UV renormalisation. Hence, for the sake of simplicity, we consider processes that are IR and UV safe but might still exhibit a local UV behaviour. For the latter, it is known that UV finite integrals might be locally divergent in the high-energy region \cite{Driencourt-Mangin:2017gop}. Therefore, a careful treatment in the UV has to be performed. For instance, at one-loop level, we refer the reader to Refs.~\cite{Becker:2010ng,Becker:2012aqa,Hernandez-Pinto:2015ysa,Sborlini:2016gbr,Sborlini:2016hat} (and references therein) and to Ref. \cite{Driencourt-Mangin:2019aix} beyond one loop. We remark that the idea of performing a local UV renormalisation is to obtain well-defined integrands in four space-time dimensions that allow a straightforward numerical evaluation.

The paper is organised as follows. In Sec.~\ref{sec:LTDintro}, we recall the basis of the LTD formalism, stressing on the formulae applied in this work. 
We briefly present its extension to the multi-loop case and a discussion regarding the treatment of loop amplitudes with multiple powers of the propagators. In Sec.~\ref{sec:kin}, we provide a description of the generation of kinematical variables by using the spinor-helicity formalism. In Sec. \ref{ssec:loop}, we focus on the parametrisation of the loop three-momenta to integrate the dual contributions when external momenta are complex. 
The introduction of local UV renormalisation counter-terms is reviewed in Sec. \ref{sec:uv}. 
The main part of this manuscript is presented in Sec.~\ref{sec:applications}, where we show numerical results for explicit examples at one-loop level. We also increase the complexity and present numerical results for $H \to gg$ at two loops in Sec.~\ref{sec:multiapp}, using the computational tools developed throughout this paper. This allows us to demonstrate the feasibility of the LTD-based numerical strategy with realistic scattering processes, as well as its efficiency. Conclusions and future research directions are analysed in Sec.~\ref{sec:conclusions}.


\section{Loop-tree duality in a nutshell}
\label{sec:LTDintro}
The loop-tree duality (LTD) theorem \cite{Catani:2008xa,Bierenbaum:2010cy,Verdugo:2020kzh,Aguilera-Verdugo:2020kzc,Ramirez-Uribe:2020hes,Aguilera-Verdugo:2020nrp} rewrites any loop integral in terms of tree-level-like that correspond to cutting, i.e. setting on-shell, internal particles. It relies on a suitable application of the Cauchy's residue theorem to reduce one-degree of freedom for each loop. We usually apply it on the energy component of the loop momenta, which translates into reducing the original Minkowski integration domain into the Euclidean space of the loop three-momenta, although it could be used to remove any other component of the loop momenta. 

In order to explain the formalism, let us consider a generic $L$-loop $N$-particle scattering amplitude, where the external momenta are labeled as $p_i$ with $i \in \{1,\ldots,N \}$. For each loop, we have an independent primitive integration variable, $\{\ell_j\}_{j=1,\ldots,L}$, and the momenta associated to the different internal lines can be written as $q_{i_s}=\ell_{s}+k_{i_s}$, where $k_{i_s}$ is a linear combination of external momenta. All the propagators depending on the same combination of primitive loop momenta, $\ell_{s}$, are enclosed together inside the set $s$. With this notation, a generic amplitude is given by
\beq
{\cal A}_N^{(L)}(1,\ldots,n) = \int_{\ell_1,\ldots,\ell_L} \sum {\cal N} \times G_F(1,\ldots,n) \, ,
\label{eq:generalAMplitud}
\eeq
where $\cal N$ represents an arbitrary numerator and 
\beq
G_F(1,\ldots,n) = \prod_{j\in 1 \cup \ldots \cup n} (G_F(q_j))^{\alpha_j} \, ,
\eeq
is a product of Feynman propagators spanned over all the possible momenta sets. Each scalar Feynman propagator can be written as
\beq  
G_F(q_i) = \frac{1}{q_i^2-m_i^2+\imath 0} = \frac{1}{q_{i,0}^2-(q_{i,0}^{(+)})^2} \, ,
\label{eq:FeynProp}
\eeq
where $q_{i,0}^{(+)} = \sqrt{\boldmath{q}_i^2+m_i^2-\imath 0}$ corresponds to the positive on-shell energy of the associated internal particle. In the previous formula, we also take into account the possibility of arbitrary powers of the propagators, through the parameters $\{\alpha_j\}$. Regarding the integration measure, we have
\beq
\int_{\ell} \equiv -\imath \, \mu^{4-d} \int \frac{d^d \ell}{(2\pi)^d} \, ,
\eeq
with $\mu$ an arbitrary energy scale to restore the proper units after the extension to a $d$-dimensional space-time. 

The LTD representation is obtained by defining the nested residues on the energy component of the propagator's momenta, $q_{i,0}$, and closing the integration contour in the lower part of the complex plane. Explicitly, if $d{\cal A}^{(L)}_F$ represents the integrand of Eq. (\ref{eq:generalAMplitud}) in the Feynman representation, the first iteration of Cauchy's theorem leads to
\beq 
{\cal A}^{(L)}_D(1;2,\ldots,n) = \sum_{i_1 \in 1} {\rm Res}(d{\cal A}_F^{(L)},{\rm Im}(q_{i_1,0})<0) \, ,
\eeq 
where we sum over all the possible configurations containing one on-shell propagator of the first set, $1$, whilst the remaining sets are left off-shell. After the $r$-th iteration, we end up with
\beqn 
&& {\cal A}^{(L)}_D(1,\ldots,r;r+1,\ldots,n) =
\\ \nonumber && \sum_{i_r \in r} {\rm Res}({\cal A}_D^{(L)}(1,\ldots,r-1;r,\ldots,n),{\rm Im}(q_{i_r,0})<0) \, ,
\eeqn 
where all the propagators in the sets to the right of the semicolon are left off-shell, and we set exactly one propagator on-shell in each one of the first $r$ sets. Once we apply the Cauchy's theorem in one of the primitive momenta, the integration measure turns into
\beq
\int_{\ell} \ \to \ \int_{\vec{\ell}} \equiv -\mu^{d-4} \int \frac{d^{d-1}\ell}{(2\pi)^{d-1}} \ , 
\eeq
i.e. transforming the $d$-dimensional Minkowski space into a $d-1$-dimensional Euclidean one. The dual LTD representation is obtained after the $L$-th iteration of the residue computation, i.e.
\beqn 
&& {\cal A}^{(L)}_N(1,\ldots,n) = 
\label{eq:masterLTD}
\\ \nonumber && \int_{\vec{\ell}_1 \ldots \vec{\ell}_L} \sum_\sigma \, {\cal A}^{(L)}_D(\sigma_1, \ldots,\sigma_L;\sigma_{L+1},\ldots,\sigma_n) \ ,
\eeqn 
where we sum over all the possible combinations of simultaneous $L$-cuts in different momenta sets. This is equivalent to perform as many cuts as loops, in order to open the loop amplitude into a set of non-disjoint trees \cite{Catani:2008xa}.

At one-loop, the beforehand mentioned formulae reduce to the usual LTD representation given in Refs. \cite{Catani:2008xa,Bierenbaum:2010cy}. It can be easily obtained by summing over all the possible single cuts, and replacing the Feynman propagators by the so-called dual propagators, namely 
\beq
G_D(q_i;q_j)= \frac{1}{q_j^2 -m_j^2 - \imath 0 \eta \cdot k_{ji}} \, ,
\label{eq:dualprop}
\eeq
where $q_j$ is the momenta flowing through the line, $q_i$ corresponds to the one that is set on-shell and $k_{ji} =q_j-q_i$. It is important to notice that the dual prescription accounts for the information contained in the multiple cuts defined within the Feynman Tree theorem (FTT)~\cite{Feynman:1963ax}. Also, that within the representation introduced in Refs. \cite{Verdugo:2020kzh,Aguilera-Verdugo:2020kzc,Ramirez-Uribe:2020hes,Aguilera-Verdugo:2020nrp}, there is no need to explicitly deal with the complex prescription of the dual propagators, since this operation is encoded within the definition of the nested residues. 

\begin{figure}[ht]
\begin{center}
\includegraphics[width=0.445\textwidth]{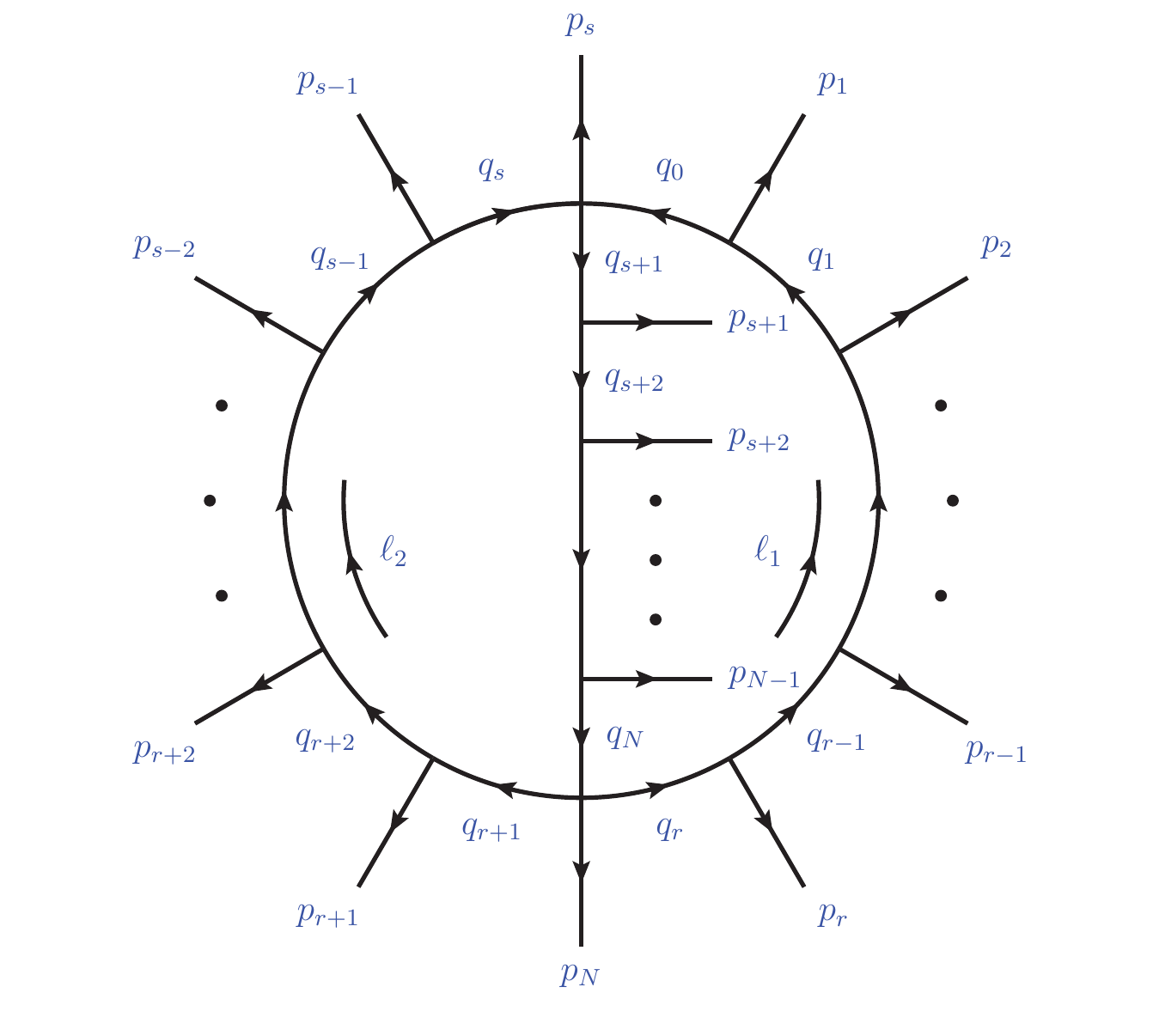} 
\caption{Diagrammatic representation of a generic two-loop diagram with $N$ external particles. The external particles can be attached to any internal line, thus defining the different sets.
\label{fig:twoloopdibujo}}
\end{center}
\end{figure}

In this article, we present practical applications up to the two-loop level, although the underlying algorithms can be extended to any loop order. In the particular two-loop case, we have $\{\ell_1,\ell_2\}$ and the sets of propagators $s \in \{1,2,3\}$, 
as shown in Fig.~\ref{fig:twoloopdibujo}. More details about the underlying subtleties of the two-loop case are available in Refs. \cite{Bierenbaum:2010cy,Driencourt-Mangin:2019aix}.

\subsection{Multiple poles and IBPs}
\label{ssec:multiplepoles}
Multi-loop integrals and local UV counter-terms can contribute with multiple powers of the propagators. 
Using integration-by-parts identities (IBPs)~\cite{Tkachov:1981wb,Chetyrkin:1981qh,Laporta:2001dd}, 
they can be reduced to linear combinations of other integrals containing only single powers of the propagators,
as done in~\cite{Bierenbaum:2012th}. 
However, this modifies the local behaviour of the integrands and might spoil the point-by-point cancellation of IR singularities present in the real-emission contribution. Thus, we will stick to the local approach and avoid using IBPs. 

The master formula given in Eq. \ref{eq:masterLTD} handles also amplitudes with multiple powers of the propagators, since its definition relies directly on the nested application of the Cauchy's theorem. A careful discussion about the computation of the residue is presented in Refs. \cite{Catani:2008xa,Bierenbaum:2012th}. It is important to take care of the dual prescription for the contributions associated to the original amplitude, since it may contain thresholds in the low energy region. In that case, the propagators associated to off-shell propagators must be promoted to dual propagators. On the contrary, when applying the LTD formalism to the UV counter-terms, we can neglect the complex prescriptions and straightforwardly use the Cauchy's formula for computing the residue.

It is worth noticing that compact formulae to obtain the LTD representation with higher-powers are presented in Ref.~\cite{Aguilera-Verdugo:2020kzc}. These expressions are obtained considering the derivatives w.r.t. $(q_{i,0}^{(+)})^2$, and taking advantage of Eq. (\ref{eq:FeynProp}) to express propagators in terms of on-shell energies.

\section{Generation of the kinematics}
\label{sec:kin}
In order to provide helicity amplitudes, we take advantage of the momentum twistor parametrisation proposed in Ref.~\cite{Hodges:2009hk}, where the standard spinor products, $\langle\bullet\,\bullet\rangle,\left[\bullet\,\bullet\right]$, are replaced by a minimal set of independent variables, $z_{i}$.
The number of variables in the latter depends on the kinematic process. In particular, any $n$-point massless amplitude can be expressed in terms of $3n-10$ independent variables. Hence, the extension to amplitudes with massive particles is straightforward. In Appendix~\ref{app:twistors}, we briefly recall the main features of these variables. 

Within the LTD approach, the evaluation of integrals is performed in the momentum space instead of using Feynman parameters or, equivalently, Mellin transformations. Then, the most suitable way of computing helicity amplitudes is through the form factors' decomposition, which has been applied within the LTD framework in Refs.~\cite{Driencourt-Mangin:2017gop,Driencourt-Mangin:2019aix}. Very recently, some alternative methods to bypass this decomposition were proposed ~\cite{Chen:2019wyb,Peraro:2019cjj}. Since the representation of the polarisation vectors may be an obstacle depending on the regularisation scheme being applied, we use the one in which external wave functions are kept in four dimensions, i.e. t'Hooft-Veltman (HV) \cite{tHooft:1972tcz} and four-dimensional helicity (FDH) \cite{Bern:2002zk} \footnote{The FDH scheme at one-loop level has been considered in a purely four-dimension formulation in Ref.~\cite{Fazio:2014xea,Mastrolia:2015maa,Primo:2016omk}.}. 

The use of HV and FDH allows us to project objects in $d$ dimensions into a four-dimensional space by making use of the following properties\footnote{We closely follow the convention of Ref.~\cite{Gnendiger:2017pys}.}:
\begin{subequations}
\begin{align}
q_{i,\left[d\right]}\cdot p_{j,\left[4\right]}&=q_{i,\left[4\right]}\cdot p_{j,\left[4\right]}\,, \\ 
 q_{i,\left[d\right]}\cdot\varepsilon_{j,[4]}&=q_{i,\left[4\right]}\cdot\varepsilon_{j,[4]}\,,
\end{align}
\end{subequations}
where we contracted the loop momentum with external momenta or polarisation vectors. Therefore, we only need to keep track of squared loop momenta, $q_{i,\left[d\right]}\cdot q_{j,\left[d\right]}$. It turns out that due to the cuts performed within the LTD formalism, we can easily remove this dependence and work with objects in four dimensions. Let us also remark that, within this approach, we do not need to include extra-dimensional products, i.e. $q_{i,\left[d-4\right]}\cdot q_{j,\left[d-4\right]}$.

Then, working in four space-time dimensions, we can parametrise the loop momenta in terms of a 
four-dimensional basis, i.e. $\mathcal{E} = \{e_i\}$. Therefore, to reduce as much as possible the number of scalar products to be evaluated, we choose $\mathcal{E} = \{p_1,p_2,\varepsilon_{12},\varepsilon_{21}\}$. With this choice, the loop momenta is expressed as
\begin{align}
q_{i}^{\alpha}&=x_{i,1}\,p_{1}^{\alpha}+x_{i,2}\,p_{2}^{\alpha}+x_{i,3}\,\varepsilon_{12}^{\alpha}+x_{i,4}\,\varepsilon_{21}^{\alpha}\,,
\end{align}
where $p_1$ and $p_2$ are the massless momenta built from the parametrisation obtained from the momentum twistors of an $n$-point kinematics and $\varepsilon_{ij}^{\alpha}=\frac{1}{2}\langle i |\gamma^\alpha|j]$. We remark that for the elements of the basis, we explicitly work with the components of the four-vectors. Hence, with this decomposition, all the scalar products involving external momenta or polarisation vectors contracted with the loop momenta can always be expressed in terms of scalar products among the elements of the basis and the loop momenta, i.e. $q_i\cdot e_j$. The aim of this refinement is twofold. Firstly, to reduce the number of scalar products required for the computation. In second place, to cancel redundant expressions that appear at integrand level. This prevents some non-contributing terms that pop up in intermediate steps of the computation, before performing an explicit evaluation.

\subsection{Parametrisation of the loop momentum}
\label{ssec:loop}
As discussed in Sec.~\ref{sec:LTDintro}, once LTD is applied to any loop integral or virtual amplitude, the integration over the loop energy component is removed and the remaining one is performed over an Euclidean space. Thus, the loop three-momentum needs to be properly parametrised to improve the computational efficiency. We remark that we are considering a complex-valued parametrisation of the external momenta. 
Explicitly, the second component of the three-momentum is purely imaginary; 
this is due to the method applied to build their representation starting from scalar invariants\footnote{For more details about the construction of the momentum parametrisation, see Appendix~\ref{app:twistors}.}. It is worth noticing, however, that the scalar products among themselves do not contain any complex phase (i.e. they are purely real), as expected in any physical kinematic configuration. Hence, to overcome any possible issue when using a real parametrisation of the loop three-momentum, we express it in cylindrical coordinates, 
\begin{align}
\boldsymbol{q}_{i} & =\left(\xi_{i}\cos\phi_{i},\rho_{i},\xi_{i}\sin\phi_{i}\right)\,,\label{eq:3momCC}
\end{align}
for the $i$-th cut. Then, the resulting integral is given by
\begin{align}
I_i & =\int_{0}^{\infty}\int_{0}^{2\pi}\int_{-\infty}^{\infty}\xi_{i}\,d\xi_{i}\,d\phi_{i}\,d\rho_{i}\,\mathcal{I}_i\left(\xi_{i},\phi_{i},\rho_{i}\right)\,,\label{eq:I1CC}
\end{align}
where $\mathcal{I}_i$ is the integrand after plugging the explicit parametrisation of the loop three-momentum~(\ref{eq:3momCC}). We note that carefully integrating $\mathcal{I}_i$ over $\rho_i$ brings large cancellations, in particular, when considering kinematical configurations below threshold. This is because the imaginary part introduced by the prescriptions must cancel in these configurations, and the $\rho_i$ variable captures all the imaginary contributions due to the explicit functional form of the  parametrisation external momenta. 
Of course, we are excluding from this claim the presence of imaginary terms introduced by the numerators (for instance, originated by the polarisation vectors). Hence, to account for the simplifications that occur in the $\rho_i$-integration, we re-write Eq.~(\ref{eq:I1CC}) as,
\begin{align}
I_i  =\int_{0}^{\infty}\int_{0}^{2\pi}&\int_{0}^{\infty}\xi_{i}\,d\xi_{i}\,d\phi_{i}\,d\rho_{i}\,\nn\\
&\times\left[\mathcal{I}_i\left(\xi_{i},\phi_{i},\rho_{i}\right)+\mathcal{I}_i\left(\xi_{i},\phi_{i},-\rho_{i}\right)\right]\, ,
\end{align}
which turns out to be equivalent to consider the real-part of the integrand in the previously mentioned conditions. 

Furthermore, we notice that the $(\xi_{i},\rho_{i})-$plane can be compactified by changing variables and using polar coordinates. Explicitly, we define
\begin{align}
\left(\xi_{i},\rho_{i}\right) & \to \frac{x_{i}}{1-x_{i}} \, \left(\cos\theta_i,\sin\theta_i\right)\,,
\end{align}
with $0\leq x_{i}<1$ and $0\leq\theta_{i}<\pi/2$. In the last part, we restricted the angular integration to the first quadrant because both $\xi_i$ and $\rho_i$ are positive.


\section{Local UV renormalisation}
\label{sec:uv}
Since we are aiming for a complete numerical implementation, it is necessary to  build integrand-level counterterms, in order to cancel the local singular behaviour at very high energies of the amplitudes under consideration. In the following, we recall how to generate these counter-terms very easily from the original amplitudes~\cite{Driencourt-Mangin:2017gop,Driencourt-Mangin:2019aix}. 

For a given loop momentum $\ell_j$, we consider the integrand-level replacement
\begin{equation}\label{UVrep}
\mathcal{S}_{j,\uv}:\{\ell_j^2~|~\ell_j\cdot k_i\}\to\{\lambda^2\,q_{j,\uv}^2+(1-\lambda^2)\mu_\uv^2~|~\lambda\,q_{j,\uv}\cdot k_i\}\,,
\end{equation}
where $\mu_\uv$ is an arbitrary scale that can be identified with the renormalisation scale, and $q_{j,\uv}=\ell_j+k_{j,\uv}$, with $k_{j,\uv}$ arbitrary, that we will set to 0 for simplicity. By applying $\mathcal{S}_{j,\uv}$ to an unintegrated and uncut one-loop amplitude $\mathcal{A}^{(1)}$ with loop momentum $\ell_j$, and then expanding in $\lambda$ around infinity up to logarithmic degree (this operation will be represented by the operator $L_\lambda$ in the following), we directly obtain an integrand-level expression that cancels the local UV singularities\footnote{This procedure is equivalent to expanding around the UV propagator $\big(G_F(q_\uv)\big)^{-1}=q_\uv^2-\mu_\uv^2+i0$ and then keeping only the divergent terms~\cite{Becker:2010ng,Becker:2012aqa,Sborlini:2016gbr,Sborlini:2016hat}.} exhibited by $\mathcal{A}^{(1)}$. It is important to note, though, that this counter-term may generate a finite part after integration, which must be fixed through a scheme fixing parameter $d_{j,\uv}$. Therefore, the counter-term reads,
\begin{equation}\label{CTj}
\mathcal{A}_{j,\uv}^{(1)}=L_\lambda\left(\mathcal{A}^{(1)}\Big|_{\mathcal{S}_{j,\uv}}\right)-d_{j,\uv}\,\mu_\uv^2\,\int_{\ell_j}\big(G_F(q_{j,\uv})\big)^3\,,
\end{equation}
where the integral multiplying $d_{j,\uv}$ integrates to the same finite quantity in both $4$ and $d$ dimensions. The quantity $\mathcal{A}_{j,\uv}^{(1)}$ properly cancels the UV behaviour of $\mathcal{A}^{(1)}$ while giving the required finite part (which is 0 for instance in the $\overline{\text{MS}}$ scheme).

For a two-loop amplitude $\mathcal{A}^{(2)}$, the general local renormalisation procedure has been extended in Ref. \cite{Driencourt-Mangin:2019aix}. In the two-loop case, it is necessary to consider three UV regimes involving the two internal momenta, $\ell_1$ and $\ell_2$. For instance, we can consider the regimes
\begin{align}\label{UVRegimes}
\begin{cases}
|\boldsymbol{\ell}_1|\to\infty\\
|\boldsymbol{\ell}_2|~~\text{fixed}
\end{cases}\!\!\!\!\!,\qquad
\begin{cases}
|\boldsymbol{\ell}_1|~~\text{fixed}\\
|\boldsymbol{\ell}_2|\to\infty
\end{cases}\!\!\!\!\!,\qquad
\begin{cases}
|\boldsymbol{\ell}_1|\to\infty\\
|\boldsymbol{\ell}_2|\to\infty
\end{cases}\!\!\!\!\!.
\end{align}
The counter-terms relative to the singular behaviour of the first two regimes can be generated using the replacement in~\Eq{UVrep}. To build the local counter-term needed to cancel the third regime, we need the additional replacement
\begin{align}
\nn\mathcal{S}_{\uv^{2}}:&\\
\nn\ell_{j}^{2}&\to\lambda^{2}\,q_{j,\uv}^{2}+(1-\lambda^{2})\mu_{\uv}^{2}\,,\\
\nn\ell_{j}\cdot\ell_{k}&\to\lambda^{2}\,q_{j,\uv}\cdot q_{k,\uv}+(1-\lambda^{2})\mu_{\uv}^{2}/2\,,\\
\ell_{j}\cdot k_{i}&\to\lambda\,q_{j,\uv}\cdot k_{i}\,,
\label{UVrep2}
\end{align}
to build the counter-term
\begin{align}\label{CT2}
\mathcal{A}_{\uv^2}=&~L_\lambda\left(\left.\left(\mathcal{A}-\,\sum_{j=1,2}\,\mathcal{A}_{j,\uv}\right)\right|_{S_{\uv^2}}\right)\nn\\
&-d_{\uv^2}\,\mu_\uv^4\int_{\ell_1}\,\int_{\ell_2}\,\big(G_F(q_{1,\uv})\big)^3\big(G_F(q_{12,\uv})\big)^3\,,
\end{align}
where once again, the term proportional to the scheme-fixing coefficient $d_{\uv^2}$ integrates to a finite quantity. 

The one- and two-loop versions of this algorithm were explicitly implemented in a {\sc{Mathematica}} code \cite{Driencourt-Mangin:2019aix}. It is fully process-independent and can be directly applied to any scattering amplitude, producing the appropriate local counter-term to regularise the divergent behaviour in the high-energy region. 


\section{Applications at one-loop}
\label{sec:applications}
In this section, we give some explicit examples in which the techniques described in Secs.~\ref{sec:kin} and~\ref{sec:uv} are applied. We focus on processes that contain two to four kinematic invariants, and we consider the non-vanishing helicity configurations. We summarise the description of our examples in Table~\ref{tab:allproc}. In the following, the kinematic invariants are implicitly given in GeV$^2$.

Since we are aiming at a calculation performed purely in four space-time dimensions, 
we restrict the analysis presented in this article to processes that are simultaneously IR and UV finite. Although the processes under consideration exhibit these features, they might still posses a local UV-divergent behaviour that prevents to perform the calculation directly in four space-time dimensions, without introducing any additional regularisation. This is because, in the most general case, the associated integrands turn out to be non-integrable functions in the high-energy limit (or UV limit).

Eventually, in the context of dimensional regularisation (DREG), setting $d=4$ from the beginning of the calculation can generate wrong results. This situation was exhaustively discussed in Ref. \cite{Driencourt-Mangin:2017gop} for the computation of the decay width of $H\to\gamma\gamma$ at leading order. Therefore, we need to build local UV counter-terms that take care of the singularities that appear at integrand level in the UV limit. In other words, we need to locally renormalise our amplitude, as explained in Sec. \ref{sec:uv}, to render the expressions integrable in four space-time dimensions.

\begin{table}[ht]
\centering
\begin{tabular}{|c|c|c|}
\hline 
$\text{Process}$ & $\begin{array}{c}
\text{Kinematic}\\
\text{scales}
\end{array}$ & $\begin{array}{c}
\text{Helicity}\\
\text{configuration}
\end{array}$\tabularnewline
\hline 
\hline 
$H\to\gamma\gamma$ & $s_{12},m_{f}^{2}$ & $++$\tabularnewline
\hline 
$\gamma\gamma\to\gamma\gamma$ & $s,t,m_{f}^{2}$ & $\begin{array}{c}
++++\\
-+++\\
--++
\end{array}$\tabularnewline
\hline 
$H\to ggg$ & $s_{12},s_{13},s_{23},m_{f}^{2}$ & $\begin{array}{c}
+++\\
-++
\end{array}$\tabularnewline
\hline 
\end{tabular}
\caption{Processes considered at one-loop level with their kinematic scales. We indicate the non-vanishing helicity configurations.}
\label{tab:allproc}
\end{table}

\begin{figure}[ht]
\centering
\includegraphics[scale=0.57]{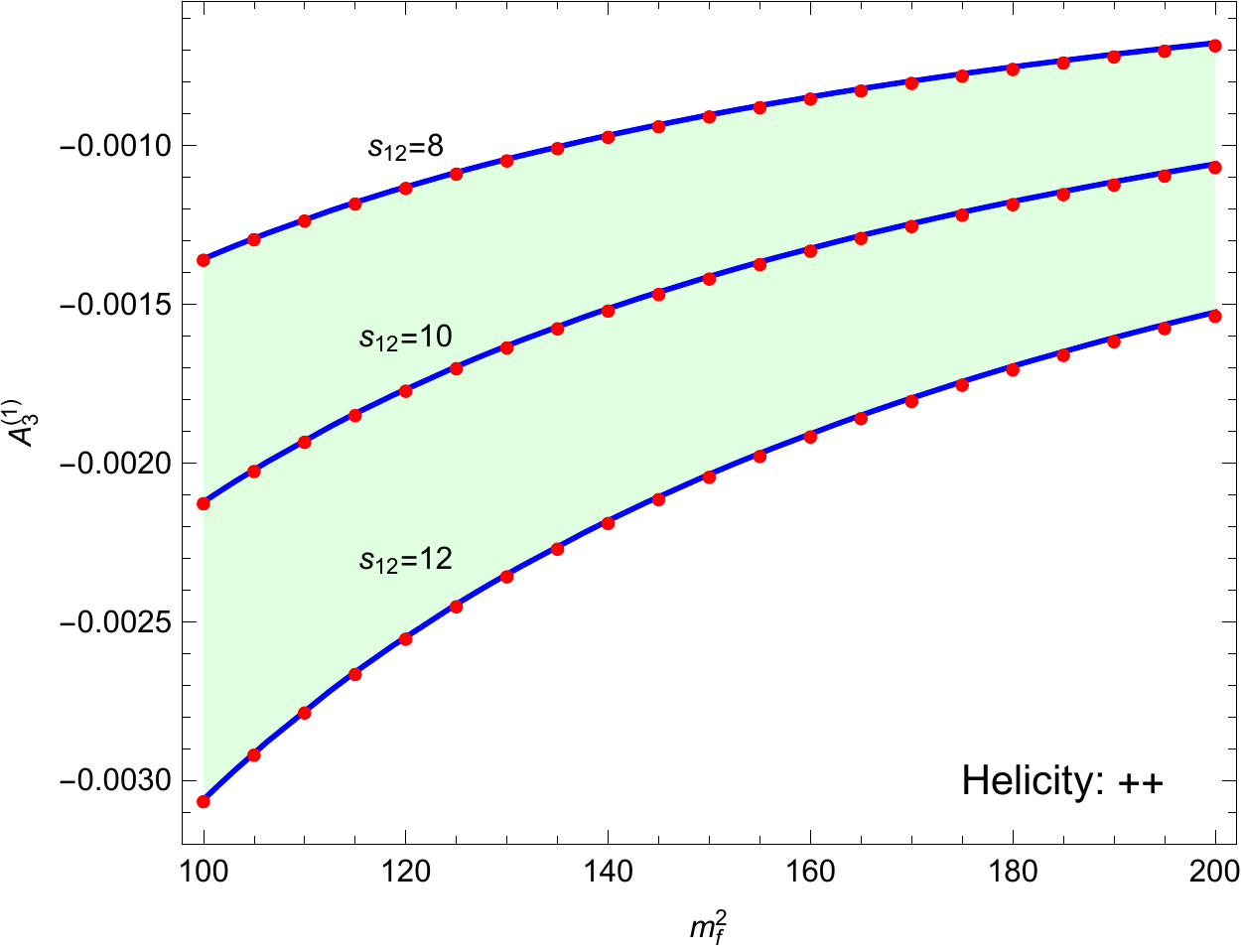} 
\caption{$H \to \gamma \gamma$ at one-loop as a function of the internal mass $m_f^2$. 
We plot the predictions for $s_{12}\in\{8,10,12\}$. The solid blue lines correspond to the analytical results, while the red points are computed through the LTD-based numerical approach.}
\label{fig:Hgamgam}
\end{figure}

The calculation of the amplitude $H\to\gamma\gamma$ performed in Ref.~\cite{Driencourt-Mangin:2017gop}, through the form factor decomposition, exploited several analytical properties in order to simplify the results. In particular, due to gauge invariance, it was possible to remove vanishing terms at integrand level. On the contrary, in the present calculation, we directly generate the proper UV counter-term to render the amplitude integrable in four space-time dimensions, without taking into account any kind of analytical property to achieve further simplifications. The numerical integration performed by LTD was compared with the analytic expression of the amplitude. For the latter, we rely on two \Mathematica\, packages, the integral reduction provided \Feyncalc~\cite{Mertig:1990an,Shtabovenko:2016sxi,Shtabovenko:2016whf} and the analytic expressions for the one-loop scalar integrals collected in~\PackageX~\cite{Patel:2015tea}. Our results are shown in Fig.~\ref{fig:Hgamgam}, where we plot the value of the amplitude as a function of the fermion internal mass, $m_f^2$, for different values of $s_{12}$. An excellent agreement is found, as expected from our previous studies of this process~\cite{Driencourt-Mangin:2017gop,Driencourt-Mangin:2019aix}.

\begin{figure}[htb]
\centering
\includegraphics[scale=0.56]{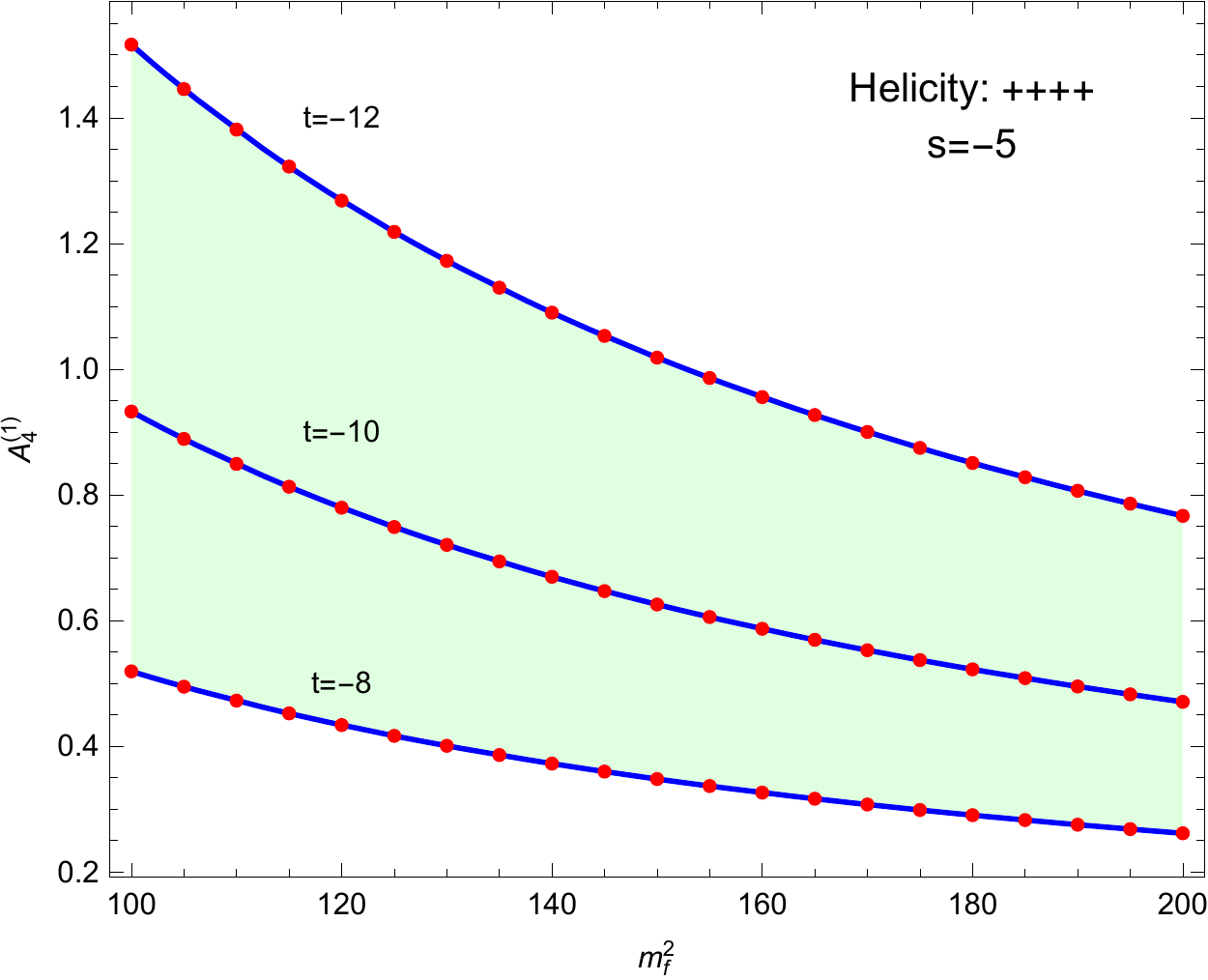}
\includegraphics[scale=0.59]{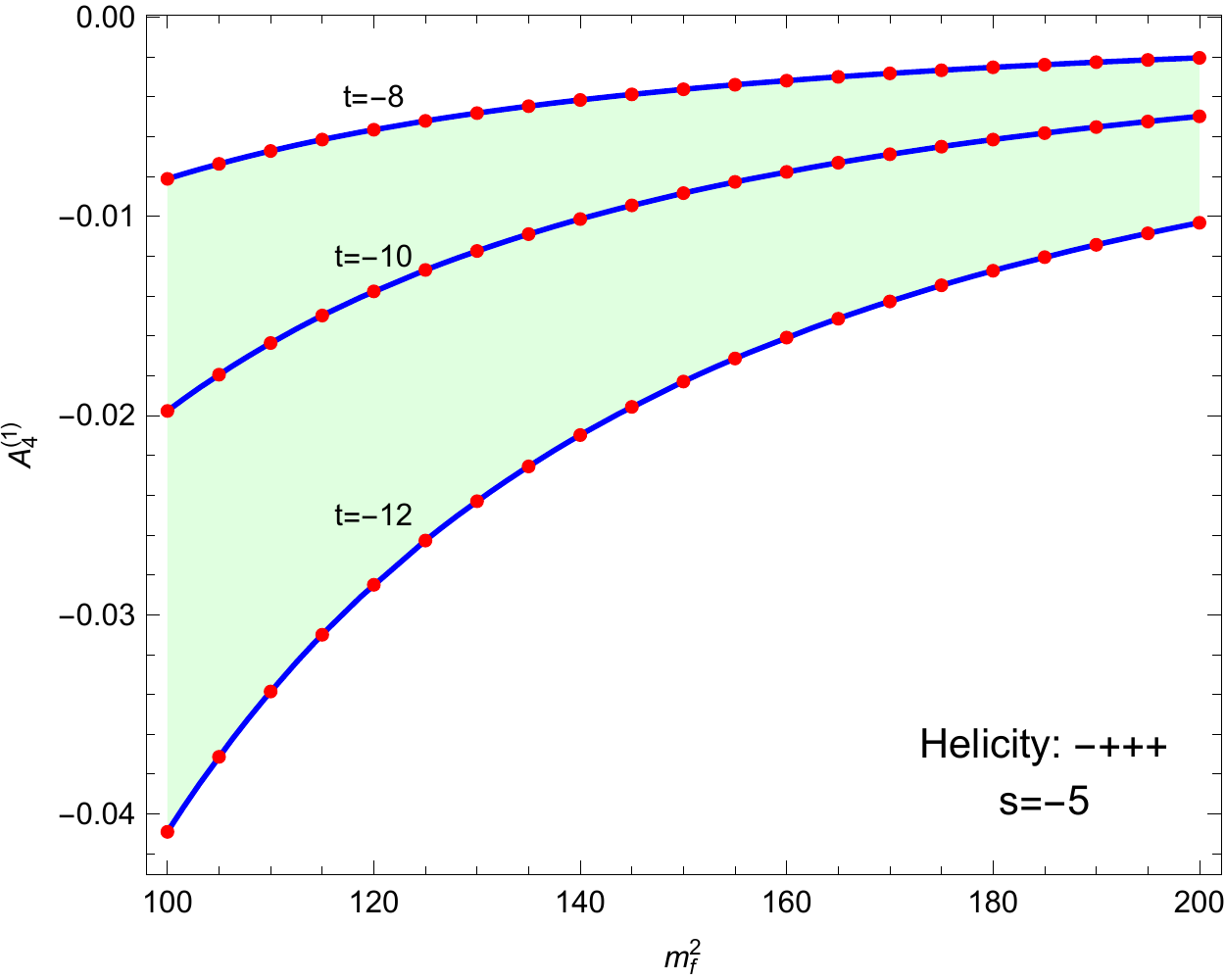}
\includegraphics[scale=0.59]{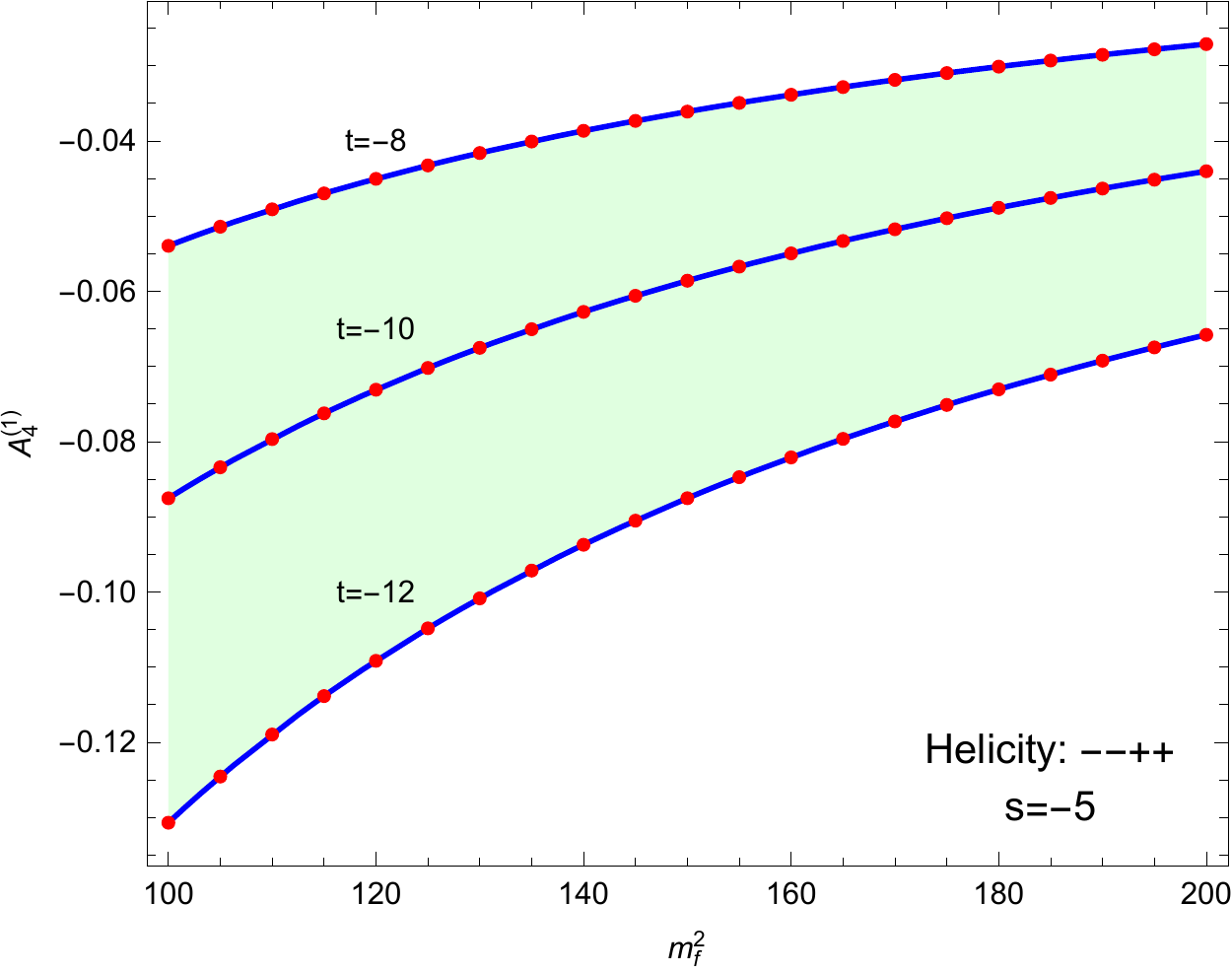}
\caption{One-loop contributions to the process $\gamma \gamma \to \gamma \gamma$, as a function of the internal mass $m_f^2$. We consider all the possible helicity configurations for a fixed ordering of the external legs: $++++$, $-+++$ and $--++$. In each case, we fix $s=-5$ and plot the predictions for $t\in\{-8,-10,-12\}$. The solid blue lines correspond to the analytical results, while the red points were computed through the LTD-based numerical approach.}
\label{fig:4photon}
\end{figure}

\begin{figure}[ht]
\centering
\includegraphics[scale=0.56]{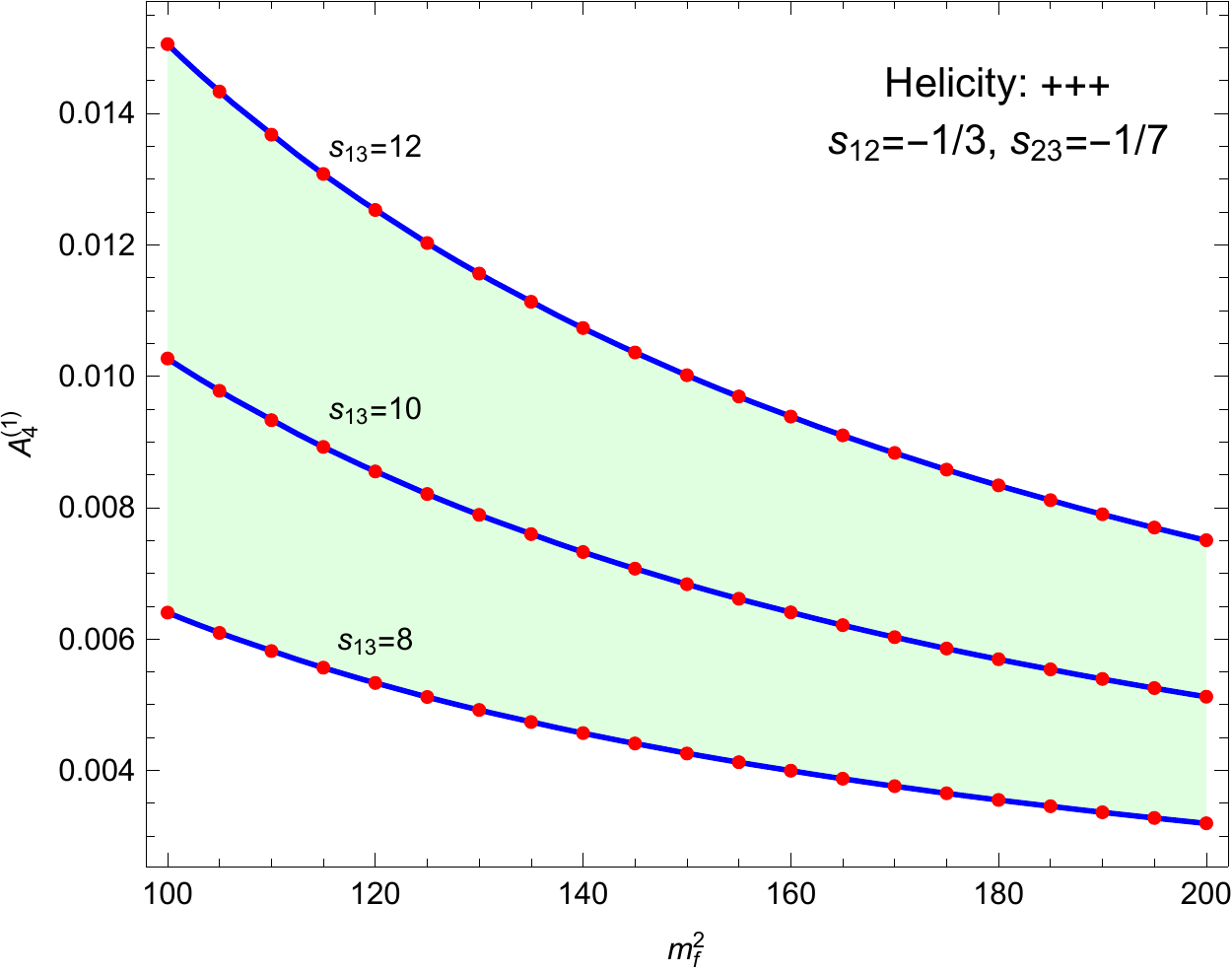} 
\includegraphics[scale=0.59]{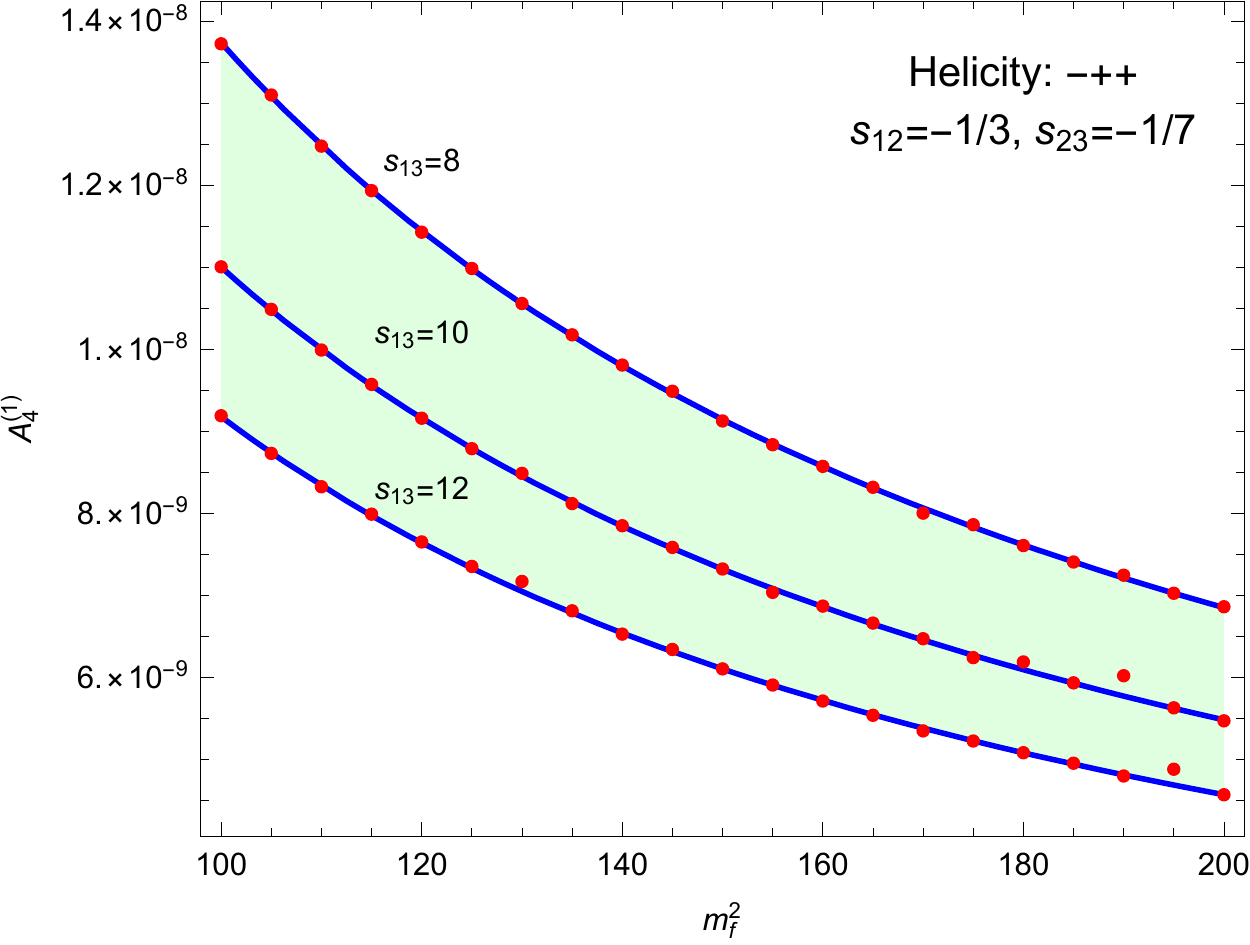}
\caption{One-loop contributions to the process $H \to g g g$, as a function of the internal mass $m_f^2$. We consider all the possible configurations for a given helicity amplitude, $+++$ and $-++$. In each case, we show the predictions for $s_{13}=\{8,10,12\}$. The solid blue lines corresponds to the analytical results, while the red points were computed through the LTD-based numerical approach.}
\label{fig:Hggg}
\end{figure}

For the processes including more kinematic scales, namely $\gamma\gamma\to\gamma\gamma$
and $H\to ggg$, we do not rely on \Feyncalc\, because it becomes inefficient when the rank of the loop momentum in the numerator starts increasing. Therefore, instead of decomposing the integrals, we work at the integrand level by reducing the amplitudes to scalar one-loop integrals. In order to do so, we follow the Ossola-Papadopoulos-Pittau (OPP) method~\cite{Ossola:2006us}
together with the integrand reduction algorithm~\cite{Mastrolia:2011pr,Badger:2012dp,Zhang:2012ce,Mastrolia:2012an,Mastrolia:2012wf,Ita:2015tya,Mastrolia:2016dhn,Mastrolia:2016czu}. 
For the evaluation of the scalar one-loop integrals we keep using \PackageX. Our results are shown in Figs.~\ref{fig:4photon} and~\ref{fig:Hggg}, where we plot the amplitudes as a function of the fermion internal mass $m_f^2$. For $\gamma \gamma \to \gamma \gamma$, we fixed $s=-5$ and considered $t=\{-8,-10,-12\}$. In the case of $H \to g g g$, $s_{12}=-1/3$ and $s_{23}=-1/7$ remained fixed while we varied $s_{13} \in [8,12]$. The agreement is very good for both processes, in all the kinematical and helicity configurations that we explored. Small numerical instabilities arise for $m_f^2>180$ in $H \to gggg$, although they can be fixed by slightly increasing the numerical precision of the integration.  

Let us stress that in the processes we consider within the LTD approach, we do not perform any integral or integrand reduction. We directly evaluate them with the proper inclusion of the UV local counter-terms, as explained in Sec.~\ref{sec:uv}. This approach indeed straightforwardly allows the evaluation of the amplitude in four space-time dimensions. Regarding the evaluation of the required integrals, we use the built-in \Mathematica\, function \verb"NIntegrate" on a desktop machine with an Intel i7 (3.4GHz) processor with 8 cores and 16 GB of RAM. The computing time for each phase-space point was $\mathcal{O}\left(30'\right)$.

\section{A two-loop example: $H \to g g$}
\label{sec:multiapp}
\begin{figure}[htb]
\centering
\includegraphics[scale=0.85]{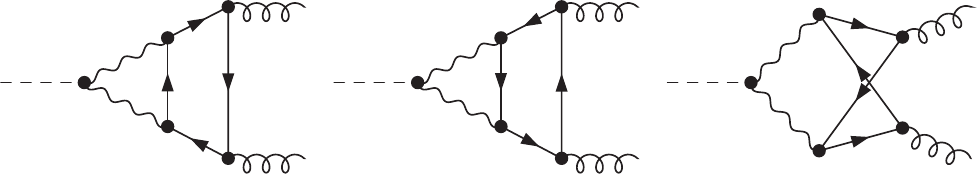}
\caption{Two-loop diagrams for the process $H\to gg$
with internal massive top quarks and $Z$ bosons.
}
\label{fig:hgg.ew-qcd}
\end{figure}

In the previous section, we demonstrated the viability of the LTD approach to tackle one-loop IR-finite amplitudes. Here, we will show that it is also a reliable strategy for two-loop processes. So, we focus on the computation of $\mathcal{O}\left(e^{3}g^{2}_S\right)$ corrections to the decay process $H \to g g$ where the internal particles are massive top quarks and $Z$ bosons. As in the examples considered at one-loop, in this case we are dealing with a finite amplitude whose contributions are given by the diagrams shown in Fig.~\ref{fig:hgg.ew-qcd}. There are not IR neither UV singularities, but it is mandatory to perform a local renormalisation. In fact, the introduction of proper local UV counter-terms is crucial to smoothly pass from $d$ to 4 dimensions when evaluating the integrals. 

In the same spirit of Ref.~\cite{Driencourt-Mangin:2019aix}, we start considering a minimal set of independent denominators. This is done in order to match the structures of the planar (P) and non-planar (NP) contributions at integrand level. Hence, we express all Feynman diagrams in terms of the following scalar integrals, 
\begin{align}
I_{\text{P/NP}} & =\int_{\ell_{1}}\int_{\ell_{2}}\frac{1}{D_{1}^{\nu_{1}}D_{2}^{\nu_{2}}D_{3}^{\nu_{3}}D_{4}^{\nu_{4}}D_{5}^{\nu_{5}}D_{6}^{\nu_{6}}D_{7}^{\nu_{7}}}\,,
\end{align}
with
\begin{align}D_{1} & =\left(\ell_{1}+\ell_{2}\right)^{2}-m_{Z}^{2}\,,\\
D_{2} & =\left(\ell_{1}+\ell_{2}+p_{1}+p_{2}\right)^{2}-m_{Z}^{2}\,,\\
D_{3} & =\ell_{1}^2-m_{t}^{2}\,,\\
D_{4} & =\left(\ell_{1}+p_{1}\right)^{2}-m_{t}^{2}\,,\\
D_{5} & =\left(\ell_{1}+p_{1}+p_{2}\right)^{2}-m_{t}^{2}\,,\\
D_{6} & =\ell_{2}^{2}-m_{t}^{2}\,,\\
D_{7} & =\left(\ell_{2}-p_{1}\right)^{2}-m_{t}^{2}\,,
\end{align}
where the auxiliary propagators are $D_{7}$ and $D_{3}$ in the planar and non-planar topologies, respectively. 

In the following, we discuss the extraction and the structure of the single and double UV local counter-terms. In particular, we have.
 
\begin{itemize}

\item Single UV counter-terms.
\begin{align}
A_{1,\text{UV}}^{\left(2\right)} & =A_{2,\text{UV}}^{\left(2\right)}=0\,.
\label{eq:1uv}
\end{align}
Let us remark that vanishing single UV counter-terms are a consequence of the way in which the propagators have been labeled. A different choice might lead to non-vanishing expressions. In order to explode this property, we labeled the propagators in such a way that the most UV-divergent contributions depend on $\ell_{1}$ and $\ell_{2}$, whilst those related to internal states depend on $\ell_{1}+\ell_{2}$ and exhibit a less UV-divergent behaviour.
\\ 
\item Double UV counter-term.

\begin{widetext}
\begin{align}
\mathcal{A}_{12,\text{UV}^{2}}^{\left(2\right)} & =2\imath\left(d-2\right)g_{f}\delta^{a_{1}a_{2}}\int_{\ell_{1}\ell_{2;}}\Bigg[-\frac{s_{12}}{D_{1\text{UV}}^{2}D_{2\text{UV}}D_{12\text{UV}}}+\frac{s_{12}}{D_{1\text{UV}}^{2}D_{12\text{UV}}^{2}}+\frac{s_{12}}{D_{1\text{UV}}D_{2\text{UV}}D_{12\text{UV}}^{2}}\nonumber \\
 & +\frac{\varepsilon_{1}\cdot q_{1}\varepsilon_{2}\cdot q_{2}}{D_{1\text{UV}}^{3}D_{2\text{UV}}D_{12\text{UV}}}-\frac{\varepsilon_{1}\cdot q_{1}\varepsilon_{2}\cdot q_{2}}{D_{1\text{UV}}^{3}D_{12\text{UV}}^{2}}-\frac{\varepsilon_{1}\cdot q_{2}\varepsilon_{2}\cdot q_{1}+2\varepsilon_{1}\cdot q_{1}\varepsilon_{2}\cdot\left(q_{1}+q_{2}\right)}{2D_{1\text{UV}}^{2}D_{2\text{UV}}D_{12\text{UV}}^{2}}\nonumber \\
 & -\frac{s_{12}}{D_{1\text{UV}}D_{2\text{UV}}D_{12\text{UV}}^{2}}-\frac{\varepsilon_{1}\cdot q_{1}\varepsilon_{2}\cdot q_{2}}{2D_{1\text{UV}}^{2}D_{2\text{UV}}^{2}D_{12\text{UV}}}+\frac{\varepsilon_{1}\cdot\left(q_{1}+q_{2}\right)\varepsilon_{2}\cdot q_{2}}{2D_{1\text{UV}}D_{2\text{UV}}^{2}D_{12\text{UV}}^{2}}+\frac{\varepsilon_{1}\cdot q_{1}\varepsilon_{2}\cdot\left(q_{1}+q_{2}\right)}{2D_{1\text{UV}}^{2}D_{2\text{UV}}D_{12\text{UV}}^{2}}\Bigg]\,,\label{eq:2uv}
\end{align}
\end{widetext}
where $\delta^{a_{1}a_{2}}$ accounts for the colour factor originated
from the two external gluons, $g_{f}=e\,g\,g'\,g_{S}^{2}\,m_{Z}\left(g_{L}^{2}+g_{R}^{2}\right)$
the overall coupling constant and the propagators in the UV limit,
$D_{i\text{UV}}=\left(q_{i}^{2}-\mu_{\text{UV}}^{2}-\imath0\right)^{-1}$,
with $i=1,2,12$. 

It is easy to check that after applying integration-by-parts identities on 
$\mathcal{A}_{12,\text{UV}^{2}}^{\left(2\right)}$, it vanishes in $d$ dimensions. In fact, as expected from the finiteness of the amplitude, the UV singularities that emerge from the planar and non-planar diagrams cancel exactly at this point. The latter is indeed in agreement with the above discussion since we are including an additional term that does not alter the behaviour in $d$ dimensions, but it collects several local features in four dimensions. In other words, we end up with an integrable function in four dimensions. 
\end{itemize}

Subtracting the counterterms in Eqs.~\eqref{eq:1uv} and~\eqref{eq:2uv} 
from the original $H \to g g$ amplitude, we tested the UV limit. 
We parametrised the loop momenta, and studied the large energy limit by performing a series expansion. All the non-integrable powers of the loop-energy cancel, 
which supports the integrability of the combined result and the local cancellation 
of the UV singularities.

\section{Conclusions and future directions}
\label{sec:conclusions}
In this article, we have explored the features of a numerical implementation based on the loop-tree duality theorem and the spinor-helicity method. We applied the dual decomposition to change the integration domain of one-loop amplitudes, from a Minkowski to an Euclidean space-time. And, simultaneously, the application of the helicity formalism lead to very compact expressions. Thus, the resulting implementation turns out to be both analytically and numerically efficient.

The computational framework developed was successfully applied to benchmark processes at one-loop. In particular, since the results were available through other codes or explicit analytical expressions, we performed a comparison and found a complete agreement with our numerical predictions. 
Additionally, we applied the two-loop local renormalisation procedure introduced in 
Ref.~\cite{Driencourt-Mangin:2019aix} in order to obtain an integrable four-dimensional representation of the $H \to g g$ helicity amplitudes with internal top quarks and $Z$ bosons.

This work constitutes an important step towards the automation of a LTD-based framework to compute physical observables in a fully numerical approach~\cite{TorresBobadilla:2019ltd}. 
By canceling the singularities directly at integrand level, we prevent them to manifest when performing the numerical evaluations and we have to deal only with integrable expressions. In this way, the algebraic handling of the loop amplitudes produces a four-dimensional representation of IR-safe finite observables. Likewise, inside our framework, we include an efficient algorithm to achieve a local renormalisation, i.e. a point-by-point cancellation of UV singularities. 
Also, taking advantage of the recent developments regarding LTD-based representations~\cite{Verdugo:2020kzh,Aguilera-Verdugo:2020kzc,Ramirez-Uribe:2020hes,Aguilera-Verdugo:2020nrp}, we expect to acquire a better understanding of the local UV renormalisation and how to increase its efficiency.


\section*{Acknowledgements}
We would like to thank Simon Badger, Pierpaolo Mastrolia, Tiziano Peraro and Amedeo Primo for illuminating discussions. This work is supported by the Spanish Government (Agencia Estatal de Investigaci\'on) and ERDF funds from European Commission (Grants No. FPA2017-84445-P and SEV-2014-0398), Generalitat Valenciana (Grant No. PROMETEO/2017/053), Consejo Superior de Investigaciones Cient\'{\i}ficas (Grant No. PIE-201750E021) and the COST Action CA16201 PARTICLEFACE. WJT acknowledges support from the Spanish Government (FJCI-2017-32128).

The Feynman diagrams depicted in this paper are generated using \Feynarts~\cite{Hahn:2000kx}.


\appendix
\section{Momentum twistor parametrisation}
\label{app:twistors}
For the sake of simplicity, we remark the main features of the twistor variables. 
Besides, for an exhaustive study of them, we refer the reader to Ref.~\cite{Badger:2016uuq,Peraro:2016wsq,TorresBobadilla:2017kpd}.

\subsection{Little group scaling}
\label{ssec:little}
Let us remark the little group scaling, a group of transformations that leaves the momentum of an on-shell particle invariant. Hence, the spinors $|i\rangle$ and $|i]$ can be re-scaled according to
\begin{align}
 & |i\rangle\to t\,|i\rangle\,, &  & |i]\to t^{-1}\,|i]\,.\label{eq:little}
\end{align}
This transformation turns out to be very interesting at the amplitude level. This is because amplitudes with massless particles can always be written in terms of spinorial products. Then, we have that: 
\begin{itemize}
\item scalar particles do not scale,
\item fermions with spin $1/2$ scale as $t^{-2h}$ for $h=\pm\frac{1}{2}$,
\item polarisation vectors with spin $1$ scale as $t^{-2h}$ for $h=\pm1$.
\end{itemize}
This implies that an $n$-point amplitude, after one of the massless particles is re-scaled according to Eq. (\ref{eq:little}), can be expressed as,
\begin{align}
&\nonumber A_{n}\left(\left\{ |1\rangle,|1],h_{1}\right\} ,\hdots,\left\{ t_{i}\,|i\rangle,t_{i}^{-1}\,|i],h_{i}\right\} ,\hdots\right)\\&=t_{i}^{-2h_{i}}\,A_{n}\left(\left\{ |1\rangle,|1],h_{1}\right\} ,\hdots,\left\{ t_{i}\,|i\rangle,t_{i}^{-1}\,|i],h_{i}\right\} ,\hdots\right)\,,
\end{align}
with $h_{i}$ the helicity of the particle $i$. 

\subsection{Momentum twistor variables}
\label{ssec:twistors}
The momentum conservation rule implies that the vectors representing the different momenta close into a contour, which can defined by the edges or by the cusps. The former is the usual representation, $p_{1}+p_{2}+\hdots+p_{n}=0$, whereas, the latter correspond to locate a point $y_{i}^{\mu}$ in
a dual space. In fact, these points can be expressed in terms of momentum vectors
\begin{align}
p_{i}^{\alpha} & =\left(y_{i}-y_{i+1}\right)^{\alpha}\,.
\end{align}
These dual variables satisfy momentum conservation after imposing a periodicity relation, namely $y_{n+1}=y_{1}$. For the sake of simplicity, we take
into account the ordering of the external particles. Hence, we define
\begin{align}
y_{ij}^{\alpha} & =\left(y_{i}-y_{j}\right)^{\alpha}=\left(p_{i}+p_{i+1}+\hdots+p_{j-1}\right)^{\alpha}\,.
\end{align}
Furthermore, because all the particles are massless (i.e $p_{i}^{2}=0$), we
write the Dirac equation in terms of holomorphic spinors,
\begin{align}
\slashed p_{i}|i\rangle & =\left(\slashed y_{i}-\slashed y_{i+1}\right)|i\rangle=0\,,
\end{align}
and we define a new variable $|\mu_{i}]$, according to 
\begin{align}
|\mu_{i}] & =\slashed y_{i}|i\rangle=\slashed y_{i+1}|i\rangle\,.
\end{align}
With these two independent variables, $|i\rangle$ and $|\mu_{i}]$,
we build a new four-component spinor variable $Z_{i}$, usually called
momentum twistor. Nevertheless, the anti-holomorphic spinors
$|i]$, can be written as 
\begin{align}
[i| & =\frac{\langle i+1\,i\rangle[\mu_{i-1}|+\langle i\,i-1\rangle[\mu_{i+1}|+\langle i-1\,i-1\rangle[\mu_{i}|}{\langle i-1\,i\rangle\langle i\,i+1\rangle}\,,
\end{align}
due to the Gordon identity. Given $n$ momentum twistors, denoted $\left(Z_{1},Z_{2},\hdots,Z_{n}\right)$, they must fulfil Poincar\'e and $\mathsf{U}\left(1\right)$
symmetries, besides satisfying momentum conservation and
on-shellness. These symmetries allow us to express any $n$-point massless
amplitude in terms for $3n-10$ variables, which is the minimal quantity required\footnote{For an extensive review of the derivation of the momentum twistors, we refer the reader to Refs.\cite{Badger:2016uuq}}.

Since we are interested in parametrising the external momenta in terms of the minimal set of variables, we follow the
representation used in Ref.~\cite{Badger:2016uuq}. In particular, for a four-point
kinematics, we have
\begin{align}
Z & =\left(\begin{array}{cccc}
|1\rangle & |2\rangle & |3\rangle & |4\rangle\\
|\mu_{1}] & |\mu_{2}] & |\mu_{3}] & |\mu_{4}]
\end{array}\right)=\left(\begin{array}{cccc}
1 & 0 & \frac{1}{z_{1}} & \frac{1}{z_{1}}+\frac{1}{z_{2}}\\
0 & 1 & 1 & 1\\
0 & 0 & -1 & -1\\
0 & 0 & 0 & 1
\end{array}\right)\,,
\end{align}
where we can relate $z_{1}$ and $z_{2}$ to the kinematic invariants according to
\begin{align}
 & z_{1}=s_{12}\,, &  & z_{2}=\frac{s_{14}}{s_{12}}\,.
\end{align}
Likewise, we obtain a particular generalisation for $n\geq5$
\begin{align}
Z & =\left(\begin{array}{cccccccc}
1 & 0 & f_{1} & f_{2} & f_{3} & \cdots & f_{n-3} & f_{n-2}\\
0 & 1 & 1 & 1 & 1 & \cdots & 1 & 1\\
0 & 0 & 0 & \frac{z_{n-1}}{z_{2}} & z_{n} & \cdots & z_{2n-6} & 1\\
0 & 0 & 1 & 1 & z_{2n-5} & \cdots & z_{3n-11} & 1-\frac{z_{3n-10}}{z_{n-1}}
\end{array}\right)\,,
\end{align}
with 
\begin{align}
f_{i} & =\sum_{k=1}^{i}\frac{1}{\prod_{l=1}^{k}z_{l}}\,,
\end{align}
and
\begin{align}
z_{i} & =\begin{cases}
s_{12} & i=1\\
-\frac{\langle i\,i+1\rangle\langle i+2\,1\rangle}{\langle1\,i\rangle\langle i+1\,i+2\rangle} & i=2,\hdots,n-2\\
\frac{s_{23}}{s_{12}} & i=n-1\\
\sum_{j=2}^{i-n+4}\frac{\langle i-n+5|j|2]}{\left[12\right]\langle1\,i-n+5\rangle} & i=n,\hdots,2n-6\\
\sum_{j=2}^{i-2n+9}\frac{\langle1|\left(2+3\right)j|i-2n+10\rangle}{s_{23}\langle1\,i-2n+10\rangle} & i=2n-5,\hdots,3n\!-\!11\\
\frac{s_{123}}{s_{12}} & i=3n-10
\end{cases}
\end{align}
We remark that with this configuration of external momenta, we drop the physical phase of the amplitude, namely, the information that accounts for parity invariance.
However, it can be straightforwardly restored by using the prefactor
\begin{align}
 & \left(\frac{\langle13\rangle}{\left[12\right]\langle23\rangle}\right)^{-h_{1}}\prod_{i=2}^{n}\left(\frac{\langle1i\rangle^{2}\left[12\right]\langle23\rangle}{\langle13\rangle}\right)^{-h_{i}}\,,
\end{align}
where $h_{i}$ are the helicities of the external massless momenta.


\section{External momenta}

In this appendix we give the external momenta in terms of the kinematic
scales shown in Secs.~\ref{sec:applications} and~\ref{sec:multiapp}.

\subsection{$H\to\gamma\gamma$ and $H\to gg$}

We focus on the process 
\begin{align}
H\left(-p_{3}\right)\to & g\left(p_{1}\right)+g\left(p_{2}\right)\,,
\end{align}
with the kinematics,
\begin{align}
p_{1}^{\mu} & =\frac{1}{2}\left\{ -1,1,\imath,-1\right\} \,,\nonumber \\
p_{2}^{\mu} & =\frac{s_{12}}{2}\left\{ 0,-1,\imath,0\right\} \,,\nonumber \\
\varepsilon_{+}^{\mu}\left(p_{1}\right) & =\frac{1}{\sqrt{2}}\left\{ 1,-1,\imath,-1\right\} \,,\nonumber \\
\varepsilon_{-}^{\mu}\left(p_{1}\right) & =\frac{1}{\sqrt{2}}\left\{ -1,0,0,-1\right\} \,,\nonumber \\
\varepsilon_{+}^{\mu}\left(p_{2}\right) & =\frac{s_{12}}{\sqrt{2}}\left\{ 1,0,0,1\right\} \,,\nonumber \\
\varepsilon_{-}^{\mu}\left(p_{2}\right) & =\frac{1}{\sqrt{2}s_{12}}\left\{ -1,1,-\imath,1\right\} \,.
\end{align}

\subsection{$\gamma\gamma\to\gamma\gamma$ }

We consider the light-by-light scattering,
\begin{align}
\gamma\left(-p_{1}\right)\gamma\left(-p_{2}\right)&\to\gamma\left(p_{3}\right)\gamma\left(p_{4}\right)\,,
\end{align}
with the kinematics,
\begin{align}
p_{1}^{\mu} & =\frac{1}{2}\left\{ -1,1,\imath,-1\right\} \,,\nonumber \\
p_{2}^{\mu} & =\frac{s}{2}\left\{ 0,-1,\imath,0\right\} \,,\nonumber \\
p_{3}^{\mu} & =\frac{1}{2}\left\{ st+1,s+t,\imath(t-s),1-st\right\} \,,\nonumber \\
\varepsilon_{+}^{\mu}\left(p_{1}\right) & =\frac{1}{\sqrt{2}}\left\{ 1,-1,\imath,-1\right\} \,,\nonumber \\
\varepsilon_{-}^{\mu}\left(p_{1}\right) & =\frac{1}{\sqrt{2}}\left\{ -1,0,0,-1\right\} \,,\nonumber \\
\varepsilon_{+}^{\mu}\left(p_{2}\right) & =\frac{s}{\sqrt{2}}\left\{ 1,0,0,1\right\} \,,\nonumber \\
\varepsilon_{-}^{\mu}\left(p_{2}\right) & =\frac{1}{\sqrt{2}s}\left\{ -1,1,-\imath,1\right\} \,.\nonumber \\
\varepsilon_{+}^{\mu}\left(p_{3}\right) & =\frac{s}{\sqrt{2}}\left\{ -1,-t,-\imath t,-s\right\} \,,\nonumber \\
\varepsilon_{-}^{\mu}\left(p_{3}\right) & =\frac{1}{\sqrt{2}s^{2}(t+1)}\left\{ s-1,-\left(s-1\right),\imath(s+1),-\left(s+1\right)\right\} \,,\nonumber \\
\varepsilon_{+}^{\mu}\left(p_{4}\right) & =\frac{st}{\sqrt{2}}\left\{ 0,1,\imath,0\right\} \,,\nonumber \\
\varepsilon_{-}^{\mu}\left(p_{4}\right) & =\frac{1}{\sqrt{2}s^{2}t^{2}}\{ -st+t+1,(s-1)t-1,\nn\\
&\qquad-\imath(st+t+1),st+t+1\} \,.
\end{align}

\subsection{$H\to ggg$}
We consider the Higgs decay into thee gluons, 
\begin{align}
H&\to g\left(p_{2}\right)g\left(p_{3}\right)g\left(p_{4}\right)\,,
\end{align}
with,
\begin{align}
p_{1}^{\mu} & =\frac{1}{2}\left\{ \frac{s_{12}+s_{13}}{s_{23}},1,\imath,\frac{s_{12}+s_{13}}{s_{23}}\right\} \,,\nonumber \\
p_{2}^{\mu} & =\frac{s_{12}}{2}\left\{ 0,-1,\imath,0\right\} \,,\nonumber \\
p_{3}^{\mu} & =\frac{1}{2}\left\{ s_{23}+1,\frac{s_{23}}{s_{12}}+s_{12},\imath\left(\frac{s_{23}}{s_{12}}-s_{12}\right),1-s_{23}\right\} \,,\nonumber \\
\varepsilon_{+}^{\mu}\left(p_{1}\right) & =\frac{1}{\sqrt{2}}\left\{ 1,-1,\imath,-1\right\} \,,\nonumber \\
\varepsilon_{-}^{\mu}\left(p_{1}\right) & =\frac{1}{\sqrt{2}}\left\{ -1,0,0,-1\right\} \,,\nonumber \\
\varepsilon_{+}^{\mu}\left(p_{2}\right) & =\frac{s_{12}}{\sqrt{2}}\left\{ 1,0,0,1\right\} \,,\nonumber \\
\varepsilon_{-}^{\mu}\left(p_{2}\right) & =\frac{1}{\sqrt{2}s_{12}}\left\{ -1,1,-\imath,1\right\} \,.\nonumber \\
\varepsilon_{+}^{\mu}\left(p_{3}\right) & =\frac{1}{\sqrt{2}}\{-s_{12},-s_{23},-\imath s_{23},-s_{12}\}\nonumber \\
\varepsilon_{-}^{\mu}\left(p_{3}\right) & =\frac{1}{\sqrt{2}s_{12}s_{13}s_{23}}\Big\{\nn\\
&-\left(s_{12}s_{23}+s_{12}+s_{13}\right),-\left(s_{12}(s_{12}+s_{13})+s_{23}\right),\nonumber \\
 &+\imath(s_{12}(s_{12}+s_{13})-s_{23}),-\left(s_{12}+s_{13}-s_{12}s_{23}\right)\Big\}\,.
\end{align}

\bibliographystyle{JHEP}
\bibliography{refs}

\providecommand{\href}[2]{#2}\begingroup\raggedright\begin{thebibliography}{10}

\bibitem{Dixon:1996wi}
L.~J. Dixon, \emph{{Calculating scattering amplitudes efficiently}},  in
  \emph{{Theoretical Advanced Study Institute in Elementary Particle Physics
  (TASI 95): QCD and Beyond}}, pp.~539--584, 1, 1996.
\newblock \href{http://arxiv.org/abs/hep-ph/9601359}{{\tt hep-ph/9601359}}.

\bibitem{Bern:1990ux}
Z.~Bern and D.~A. Kosower, \emph{{Color decomposition of one loop amplitudes in
  gauge theories}},
  \href{http://dx.doi.org/10.1016/0550-3213(91)90567-H}{\emph{Nucl. Phys. B}
  {\bf 362} (1991) 389--448}.

\bibitem{Kleiss:1988ne}
R.~Kleiss and H.~Kuijf, \emph{{Multi - Gluon Cross-sections and Five Jet
  Production at Hadron Colliders}},
  \href{http://dx.doi.org/10.1016/0550-3213(89)90574-9}{\emph{Nucl. Phys. B}
  {\bf 312} (1989) 616--644}.

\bibitem{Bern:2008qj}
Z.~Bern, J.~J.~M. Carrasco and H.~Johansson, \emph{{New Relations for
  Gauge-Theory Amplitudes}},
  \href{http://dx.doi.org/10.1103/PhysRevD.78.085011}{\emph{Phys. Rev. D} {\bf
  78} (2008) 085011}, [\href{http://arxiv.org/abs/0805.3993}{{\tt 0805.3993}}].

\bibitem{Gnendiger:2017pys}
C.~Gnendiger et~al., \emph{{To ${d}$, or not to ${d}$: recent developments and
  comparisons of regularization schemes}},
  \href{http://dx.doi.org/10.1140/epjc/s10052-017-5023-2}{\emph{Eur. Phys. J.
  C} {\bf 77} (2017) 471}, [\href{http://arxiv.org/abs/1705.01827}{{\tt
  1705.01827}}].

\bibitem{TorresBobadilla:2020ekr}
W.~J. Torres~Bobadilla et~al., \emph{{May the four be with you: Novel
  IR-subtraction methods to tackle NNLO calculations}},
  \href{http://arxiv.org/abs/2012.02567}{{\tt 2012.02567}}.

\bibitem{Heinrich:2020ybq}
G.~Heinrich, \emph{{Collider Physics at the Precision Frontier}},
  \href{http://arxiv.org/abs/2009.00516}{{\tt 2009.00516}}.

\bibitem{Catani:2008xa}
S.~Catani, T.~Gleisberg, F.~Krauss, G.~Rodrigo and J.-C. Winter, \emph{{From
  loops to trees by-passing Feynman's theorem}},
  \href{http://dx.doi.org/10.1088/1126-6708/2008/09/065}{\emph{JHEP} {\bf 09}
  (2008) 065}, [\href{http://arxiv.org/abs/0804.3170}{{\tt 0804.3170}}].

\bibitem{Bierenbaum:2010cy}
I.~Bierenbaum, S.~Catani, P.~Draggiotis and G.~Rodrigo, \emph{{A Tree-Loop
  Duality Relation at Two Loops and Beyond}},
  \href{http://dx.doi.org/10.1007/JHEP10(2010)073}{\emph{JHEP} {\bf 10} (2010)
  073}, [\href{http://arxiv.org/abs/1007.0194}{{\tt 1007.0194}}].

\bibitem{Bierenbaum:2012th}
I.~Bierenbaum, S.~Buchta, P.~Draggiotis, I.~Malamos and G.~Rodrigo,
  \emph{{Tree-Loop Duality Relation beyond simple poles}},
  \href{http://dx.doi.org/10.1007/JHEP03(2013)025}{\emph{JHEP} {\bf 03} (2013)
  025}, [\href{http://arxiv.org/abs/1211.5048}{{\tt 1211.5048}}].

\bibitem{Aguilera-Verdugo:2019kbz}
J.~J. Aguilera-Verdugo, F.~Driencourt-Mangin, J.~Plenter,
  S.~Ram\'\i{}rez-Uribe, G.~Rodrigo, G.~F.~R. Sborlini et~al.,
  \emph{{Causality, unitarity thresholds, anomalous thresholds and infrared
  singularities from the loop-tree duality at higher orders}},
  \href{http://dx.doi.org/10.1007/JHEP12(2019)163}{\emph{JHEP} {\bf 12} (2019)
  163}, [\href{http://arxiv.org/abs/1904.08389}{{\tt 1904.08389}}].

\bibitem{Verdugo:2020kzh}
J.~J. Aguilera-Verdugo, F.~Driencourt-Mangin, R.~J. Hern\'andez-Pinto,
  J.~Plenter, S.~Ramirez-Uribe, A.~E. Renteria~Olivo et~al., \emph{{Open Loop
  Amplitudes and Causality to All Orders and Powers from the Loop-Tree
  Duality}},
  \href{http://dx.doi.org/10.1103/PhysRevLett.124.211602}{\emph{Phys. Rev.
  Lett.} {\bf 124} (2020) 211602}, [\href{http://arxiv.org/abs/2001.03564}{{\tt
  2001.03564}}].

\bibitem{Plenter:2020lop}
J.~Plenter and G.~Rodrigo, \emph{{Asymptotic expansions through the loop-tree
  duality}},  \href{http://arxiv.org/abs/2005.02119}{{\tt 2005.02119}}.

\bibitem{Aguilera-Verdugo:2020kzc}
J.~J. Aguilera-Verdugo, R.~J. Hernandez-Pinto, G.~Rodrigo, G.~F.~R. Sborlini
  and W.~J. Torres~Bobadilla, \emph{{Causal representation of multi-loop
  amplitudes within the loop-tree duality}},
  \href{http://dx.doi.org/10.1007/JHEP01(2021)069}{\emph{JHEP} {\bf 01} (2021)
  069}, [\href{http://arxiv.org/abs/2006.11217}{{\tt 2006.11217}}].

\bibitem{Ramirez-Uribe:2020hes}
S.~Ram\'\i{}rez-Uribe, R.~J. Hern\'andez-Pinto, G.~Rodrigo, G.~F.~R. Sborlini
  and W.~J. Torres~Bobadilla, \emph{{Universal opening of four-loop scattering
  amplitudes to trees}},  \href{http://arxiv.org/abs/2006.13818}{{\tt
  2006.13818}}.

\bibitem{Aguilera-Verdugo:2020nrp}
J.~J. Aguilera-Verdugo, R.~J. Hernandez-Pinto, G.~Rodrigo, G.~F.~R. Sborlini
  and W.~J. Torres~Bobadilla, \emph{{Mathematical properties of nested residues
  and their application to multi-loop scattering amplitudes}},
  \href{http://arxiv.org/abs/2010.12971}{{\tt 2010.12971}}.

\bibitem{Tomboulis:2017rvd}
E.~T. Tomboulis, \emph{{Causality and Unitarity via the Tree-Loop Duality
  Relation}}, \href{http://dx.doi.org/10.1007/JHEP05(2017)148}{\emph{JHEP} {\bf
  05} (2017) 148}, [\href{http://arxiv.org/abs/1701.07052}{{\tt 1701.07052}}].

\bibitem{Runkel:2019yrs}
R.~Runkel, Z.~Sz\H{o}r, J.~P. Vesga and S.~Weinzierl, \emph{{Causality and
  loop-tree duality at higher loops}},
  \href{http://dx.doi.org/10.1103/PhysRevLett.122.111603}{\emph{Phys. Rev.
  Lett.} {\bf 122} (2019) 111603}, [\href{http://arxiv.org/abs/1902.02135}{{\tt
  1902.02135}}].

\bibitem{Capatti:2019ypt}
Z.~Capatti, V.~Hirschi, D.~Kermanschah and B.~Ruijl, \emph{{Loop-Tree Duality
  for Multiloop Numerical Integration}},
  \href{http://dx.doi.org/10.1103/PhysRevLett.123.151602}{\emph{Phys. Rev.
  Lett.} {\bf 123} (2019) 151602}, [\href{http://arxiv.org/abs/1906.06138}{{\tt
  1906.06138}}].

\bibitem{Capatti:2020ytd}
Z.~Capatti, V.~Hirschi, D.~Kermanschah, A.~Pelloni and B.~Ruijl,
  \emph{{Manifestly Causal Loop-Tree Duality}},
  \href{http://arxiv.org/abs/2009.05509}{{\tt 2009.05509}}.

\bibitem{Capatti:2020xjc}
Z.~Capatti, V.~Hirschi, A.~Pelloni and B.~Ruijl, \emph{{Local Unitarity: a
  representation of differential cross-sections that is locally free of
  infrared singularities at any order}},
  \href{http://arxiv.org/abs/2010.01068}{{\tt 2010.01068}}.

\bibitem{Prisco:2020kyb}
R.~M. Prisco and F.~Tramontano, \emph{{Dual Subtractions}},
  \href{http://arxiv.org/abs/2012.05012}{{\tt 2012.05012}}.

\bibitem{Hodges:2009hk}
A.~Hodges, \emph{{Eliminating spurious poles from gauge-theoretic amplitudes}},
  \href{http://dx.doi.org/10.1007/JHEP05(2013)135}{\emph{JHEP} {\bf 05} (2013)
  135}, [\href{http://arxiv.org/abs/0905.1473}{{\tt 0905.1473}}].

\bibitem{Badger:2013gxa}
S.~Badger, H.~Frellesvig and Y.~Zhang, \emph{{A Two-Loop Five-Gluon Helicity
  Amplitude in QCD}},
  \href{http://dx.doi.org/10.1007/JHEP12(2013)045}{\emph{JHEP} {\bf 12} (2013)
  045}, [\href{http://arxiv.org/abs/1310.1051}{{\tt 1310.1051}}].

\bibitem{Badger:2016uuq}
S.~Badger, \emph{{Automating QCD amplitudes with on-shell methods}},
  \href{http://dx.doi.org/10.1088/1742-6596/762/1/012057}{\emph{J. Phys. Conf.
  Ser.} {\bf 762} (2016) 012057}, [\href{http://arxiv.org/abs/1605.02172}{{\tt
  1605.02172}}].

\bibitem{Driencourt-Mangin:2017gop}
F.~Driencourt-Mangin, G.~Rodrigo and G.~F.~R. Sborlini, \emph{{Universal dual
  amplitudes and asymptotic expansions for $gg\rightarrow H$ and $H\rightarrow
  \gamma \gamma $ in four dimensions}},
  \href{http://dx.doi.org/10.1140/epjc/s10052-018-5692-5}{\emph{Eur. Phys. J.
  C} {\bf 78} (2018) 231}, [\href{http://arxiv.org/abs/1702.07581}{{\tt
  1702.07581}}].

\bibitem{Becker:2010ng}
S.~Becker, C.~Reuschle and S.~Weinzierl, \emph{{Numerical NLO QCD
  calculations}}, \href{http://dx.doi.org/10.1007/JHEP12(2010)013}{\emph{JHEP}
  {\bf 12} (2010) 013}, [\href{http://arxiv.org/abs/1010.4187}{{\tt
  1010.4187}}].

\bibitem{Becker:2012aqa}
S.~Becker, C.~Reuschle and S.~Weinzierl, \emph{{Efficiency Improvements for the
  Numerical Computation of NLO Corrections}},
  \href{http://dx.doi.org/10.1007/JHEP07(2012)090}{\emph{JHEP} {\bf 07} (2012)
  090}, [\href{http://arxiv.org/abs/1205.2096}{{\tt 1205.2096}}].

\bibitem{Hernandez-Pinto:2015ysa}
R.~J. Hernandez-Pinto, G.~F.~R. Sborlini and G.~Rodrigo, \emph{{Towards gauge
  theories in four dimensions}},
  \href{http://dx.doi.org/10.1007/JHEP02(2016)044}{\emph{JHEP} {\bf 02} (2016)
  044}, [\href{http://arxiv.org/abs/1506.04617}{{\tt 1506.04617}}].

\bibitem{Sborlini:2016gbr}
G.~F.~R. Sborlini, F.~Driencourt-Mangin, R.~Hernandez-Pinto and G.~Rodrigo,
  \emph{{Four-dimensional unsubtraction from the loop-tree duality}},
  \href{http://dx.doi.org/10.1007/JHEP08(2016)160}{\emph{JHEP} {\bf 08} (2016)
  160}, [\href{http://arxiv.org/abs/1604.06699}{{\tt 1604.06699}}].

\bibitem{Sborlini:2016hat}
G.~F.~R. Sborlini, F.~Driencourt-Mangin and G.~Rodrigo, \emph{{Four-dimensional
  unsubtraction with massive particles}},
  \href{http://dx.doi.org/10.1007/JHEP10(2016)162}{\emph{JHEP} {\bf 10} (2016)
  162}, [\href{http://arxiv.org/abs/1608.01584}{{\tt 1608.01584}}].

\bibitem{Driencourt-Mangin:2019aix}
F.~Driencourt-Mangin, G.~Rodrigo, G.~F.~R. Sborlini and W.~J. Torres~Bobadilla,
  \emph{{Universal four-dimensional representation of $H \to \gamma \gamma$ at
  two loops through the Loop-Tree Duality}},
  \href{http://dx.doi.org/10.1007/JHEP02(2019)143}{\emph{JHEP} {\bf 02} (2019)
  143}, [\href{http://arxiv.org/abs/1901.09853}{{\tt 1901.09853}}].

\bibitem{Feynman:1963ax}
R.~P. Feynman, \emph{{Quantum theory of gravitation}}, {\emph{Acta Phys.
  Polon.} {\bf 24} (1963) 697--722}.

\bibitem{Tkachov:1981wb}
F.~V. Tkachov, \emph{{A Theorem on Analytical Calculability of Four Loop
  Renormalization Group Functions}},
  \href{http://dx.doi.org/10.1016/0370-2693(81)90288-4}{\emph{Phys. Lett. B}
  {\bf 100} (1981) 65--68}.

\bibitem{Chetyrkin:1981qh}
K.~G. Chetyrkin and F.~V. Tkachov, \emph{{Integration by Parts: The Algorithm
  to Calculate beta Functions in 4 Loops}},
  \href{http://dx.doi.org/10.1016/0550-3213(81)90199-1}{\emph{Nucl. Phys. B}
  {\bf 192} (1981) 159--204}.

\bibitem{Laporta:2001dd}
S.~Laporta, \emph{{High precision calculation of multiloop Feynman integrals by
  difference equations}},
  \href{http://dx.doi.org/10.1016/S0217-751X(00)00215-7}{\emph{Int. J. Mod.
  Phys. A} {\bf 15} (2000) 5087--5159},
  [\href{http://arxiv.org/abs/hep-ph/0102033}{{\tt hep-ph/0102033}}].

\bibitem{Chen:2019wyb}
L.~Chen, \emph{{A prescription for projectors to compute helicity amplitudes in
  D dimensions}},  \href{http://arxiv.org/abs/1904.00705}{{\tt 1904.00705}}.

\bibitem{Peraro:2019cjj}
T.~Peraro and L.~Tancredi, \emph{{Physical projectors for multi-leg helicity
  amplitudes}}, \href{http://dx.doi.org/10.1007/JHEP07(2019)114}{\emph{JHEP}
  {\bf 07} (2019) 114}, [\href{http://arxiv.org/abs/1906.03298}{{\tt
  1906.03298}}].

\bibitem{tHooft:1972tcz}
G.~'t~Hooft and M.~J.~G. Veltman, \emph{{Regularization and Renormalization of
  Gauge Fields}},
  \href{http://dx.doi.org/10.1016/0550-3213(72)90279-9}{\emph{Nucl. Phys. B}
  {\bf 44} (1972) 189--213}.

\bibitem{Bern:2002zk}
Z.~Bern, A.~De~Freitas, L.~J. Dixon and H.~L. Wong, \emph{{Supersymmetric
  regularization, two loop QCD amplitudes and coupling shifts}},
  \href{http://dx.doi.org/10.1103/PhysRevD.66.085002}{\emph{Phys. Rev. D} {\bf
  66} (2002) 085002}, [\href{http://arxiv.org/abs/hep-ph/0202271}{{\tt
  hep-ph/0202271}}].

\bibitem{Fazio:2014xea}
R.~A. Fazio, P.~Mastrolia, E.~Mirabella and W.~J. Torres~Bobadilla, \emph{{On
  the Four-Dimensional Formulation of Dimensionally Regulated Amplitudes}},
  \href{http://dx.doi.org/10.1140/epjc/s10052-014-3197-4}{\emph{Eur. Phys. J.
  C} {\bf 74} (2014) 3197}, [\href{http://arxiv.org/abs/1404.4783}{{\tt
  1404.4783}}].

\bibitem{Mastrolia:2015maa}
P.~Mastrolia, A.~Primo, U.~Schubert and W.~J. Torres~Bobadilla,
  \emph{{Off-shell currents and color\textendash{}kinematics duality}},
  \href{http://dx.doi.org/10.1016/j.physletb.2015.11.084}{\emph{Phys. Lett. B}
  {\bf 753} (2016) 242--262}, [\href{http://arxiv.org/abs/1507.07532}{{\tt
  1507.07532}}].

\bibitem{Primo:2016omk}
A.~Primo and W.~J. Torres~Bobadilla, \emph{{BCJ Identities and $d$-Dimensional
  Generalized Unitarity}},
  \href{http://dx.doi.org/10.1007/JHEP04(2016)125}{\emph{JHEP} {\bf 04} (2016)
  125}, [\href{http://arxiv.org/abs/1602.03161}{{\tt 1602.03161}}].

\bibitem{Mertig:1990an}
R.~Mertig, M.~Bohm and A.~Denner, \emph{{FEYN CALC: Computer algebraic
  calculation of Feynman amplitudes}},
  \href{http://dx.doi.org/10.1016/0010-4655(91)90130-D}{\emph{Comput. Phys.
  Commun.} {\bf 64} (1991) 345--359}.

\bibitem{Shtabovenko:2016sxi}
V.~Shtabovenko, R.~Mertig and F.~Orellana, \emph{{New Developments in FeynCalc
  9.0}}, \href{http://dx.doi.org/10.1016/j.cpc.2016.06.008}{\emph{Comput. Phys.
  Commun.} {\bf 207} (2016) 432--444},
  [\href{http://arxiv.org/abs/1601.01167}{{\tt 1601.01167}}].

\bibitem{Shtabovenko:2016whf}
V.~Shtabovenko, \emph{{FeynHelpers: Connecting FeynCalc to FIRE and
  Package-X}}, \href{http://dx.doi.org/10.1016/j.cpc.2017.04.014}{\emph{Comput.
  Phys. Commun.} {\bf 218} (2017) 48--65},
  [\href{http://arxiv.org/abs/1611.06793}{{\tt 1611.06793}}].

\bibitem{Patel:2015tea}
H.~H. Patel, \emph{{Package-X: A Mathematica package for the analytic
  calculation of one-loop integrals}},
  \href{http://dx.doi.org/10.1016/j.cpc.2015.08.017}{\emph{Comput. Phys.
  Commun.} {\bf 197} (2015) 276--290},
  [\href{http://arxiv.org/abs/1503.01469}{{\tt 1503.01469}}].

\bibitem{Ossola:2006us}
G.~Ossola, C.~G. Papadopoulos and R.~Pittau, \emph{{Reducing full one-loop
  amplitudes to scalar integrals at the integrand level}},
  \href{http://dx.doi.org/10.1016/j.nuclphysb.2006.11.012}{\emph{Nucl. Phys. B}
  {\bf 763} (2007) 147--169}, [\href{http://arxiv.org/abs/hep-ph/0609007}{{\tt
  hep-ph/0609007}}].

\bibitem{Mastrolia:2011pr}
P.~Mastrolia and G.~Ossola, \emph{{On the Integrand-Reduction Method for
  Two-Loop Scattering Amplitudes}},
  \href{http://dx.doi.org/10.1007/JHEP11(2011)014}{\emph{JHEP} {\bf 11} (2011)
  014}, [\href{http://arxiv.org/abs/1107.6041}{{\tt 1107.6041}}].

\bibitem{Badger:2012dp}
S.~Badger, H.~Frellesvig and Y.~Zhang, \emph{{Hepta-Cuts of Two-Loop Scattering
  Amplitudes}}, \href{http://dx.doi.org/10.1007/JHEP04(2012)055}{\emph{JHEP}
  {\bf 04} (2012) 055}, [\href{http://arxiv.org/abs/1202.2019}{{\tt
  1202.2019}}].

\bibitem{Zhang:2012ce}
Y.~Zhang, \emph{{Integrand-Level Reduction of Loop Amplitudes by Computational
  Algebraic Geometry Methods}},
  \href{http://dx.doi.org/10.1007/JHEP09(2012)042}{\emph{JHEP} {\bf 09} (2012)
  042}, [\href{http://arxiv.org/abs/1205.5707}{{\tt 1205.5707}}].

\bibitem{Mastrolia:2012an}
P.~Mastrolia, E.~Mirabella, G.~Ossola and T.~Peraro, \emph{{Scattering
  Amplitudes from Multivariate Polynomial Division}},
  \href{http://dx.doi.org/10.1016/j.physletb.2012.09.053}{\emph{Phys. Lett. B}
  {\bf 718} (2012) 173--177}, [\href{http://arxiv.org/abs/1205.7087}{{\tt
  1205.7087}}].

\bibitem{Mastrolia:2012wf}
P.~Mastrolia, E.~Mirabella, G.~Ossola and T.~Peraro, \emph{{Integrand-Reduction
  for Two-Loop Scattering Amplitudes through Multivariate Polynomial
  Division}}, \href{http://dx.doi.org/10.1103/PhysRevD.87.085026}{\emph{Phys.
  Rev. D} {\bf 87} (2013) 085026}, [\href{http://arxiv.org/abs/1209.4319}{{\tt
  1209.4319}}].

\bibitem{Ita:2015tya}
H.~Ita, \emph{{Two-loop Integrand Decomposition into Master Integrals and
  Surface Terms}},
  \href{http://dx.doi.org/10.1103/PhysRevD.94.116015}{\emph{Phys. Rev. D} {\bf
  94} (2016) 116015}, [\href{http://arxiv.org/abs/1510.05626}{{\tt
  1510.05626}}].

\bibitem{Mastrolia:2016dhn}
P.~Mastrolia, T.~Peraro and A.~Primo, \emph{{Adaptive Integrand Decomposition
  in parallel and orthogonal space}},
  \href{http://dx.doi.org/10.1007/JHEP08(2016)164}{\emph{JHEP} {\bf 08} (2016)
  164}, [\href{http://arxiv.org/abs/1605.03157}{{\tt 1605.03157}}].

\bibitem{Mastrolia:2016czu}
P.~Mastrolia, T.~Peraro, A.~Primo and W.~J. Torres~Bobadilla, \emph{{Adaptive
  Integrand Decomposition}},
  \href{http://dx.doi.org/10.22323/1.260.0007}{\emph{PoS} {\bf LL2016} (2016)
  007}, [\href{http://arxiv.org/abs/1607.05156}{{\tt 1607.05156}}].

\bibitem{TorresBobadilla:2019ltd}
W.~J. Torres~Bobadilla, \emph{{Lotty: the LOop-Tree dualiTY automation}},
  {\emph{In preparation} }.

\bibitem{Hahn:2000kx}
T.~Hahn, \emph{{Generating Feynman diagrams and amplitudes with FeynArts 3}},
  \href{http://dx.doi.org/10.1016/S0010-4655(01)00290-9}{\emph{Comput. Phys.
  Commun.} {\bf 140} (2001) 418--431},
  [\href{http://arxiv.org/abs/hep-ph/0012260}{{\tt hep-ph/0012260}}].

\bibitem{Peraro:2016wsq}
T.~Peraro, \emph{{Scattering amplitudes over finite fields and multivariate
  functional reconstruction}},
  \href{http://dx.doi.org/10.1007/JHEP12(2016)030}{\emph{JHEP} {\bf 12} (2016)
  030}, [\href{http://arxiv.org/abs/1608.01902}{{\tt 1608.01902}}].

\bibitem{TorresBobadilla:2017kpd}
W.~J. Torres~Bobadilla, \emph{{Generalised Unitarity, Integrand Decomposition,
  and Hidden properties of QCD Scattering Amplitudes in Dimensional
  Regularisation}}.
\newblock PhD thesis, Padua U., 2017.

\end{thebibliography}\endgroup

\end{document}